\documentclass[10pt,aps,twocolumn,prl,superscriptaddress,bibnotes,amsmath,amsfonts,floatfix,nolongbibliography,a4paper]{revtex4-2}
\pdfoutput=1
\usepackage{graphicx}
\usepackage{multirow}
\usepackage{graphicx}
\usepackage{verbatim}
\usepackage{amsmath}
\usepackage{soul}

\usepackage{lmodern}
\usepackage{textcomp}

\IfFileExists{microtype.sty}{\usepackage{microtype}}{\relax}
\usepackage[pdfusetitle]{hyperref}

\def\0{\phantom{0}}

\newcommand{\I}{\mathrm{i}}
\newcommand{\beq}[1]{\begin{equation} #1 \end{equation}}
\newcommand{\bsplit}[1]{\begin{equation} \begin{split} #1 \end{split} \end{equation}}

\usepackage[dvipsnames]{xcolor}

\begin{document} 

\title{Collective modes and Raman response in Ta$_2$NiSe$_5$}
\author{Banhi Chatterjee}
\affiliation{Jožef Stefan Institute, Jamova 39, SI-1000 Ljubljana, Slovenia}
\author{Jernej Mravlje}
\affiliation{Jožef Stefan Institute, Jamova 39, SI-1000 Ljubljana, Slovenia}
\affiliation{Faculty of Mathematics and Physics, University of Ljubljana, Jadranska 19, 1000 Ljubljana, Slovenia} 
\author{Denis Gole\v z}
\affiliation{Jožef Stefan Institute, Jamova 39, SI-1000 Ljubljana, Slovenia}
\affiliation{Faculty of Mathematics and Physics, University of Ljubljana, Jadranska 19, 1000 Ljubljana, Slovenia}

\date{\today}

\begin{abstract}
We explore the collective response in an excitonic insulator phase in Ta$_2$NiSe$_5$ using a semirealistic model including relevant lattice and electronic instabilities.  We calculate order-parameter susceptibility and Raman response within a time-dependent Hartree-Fock approach.  Contrary to the standard expectations, the amplitude mode frequency does not coincide with the single-particle gap but has a higher frequency. We find a phase mode that is massive because the excitonic condensation breaks a discrete symmetry only and that becomes heavier as the electron-lattice coupling is increased.  These features are expected to apply to generic realistic excitonic insulators. 
We discuss scenarios under which the phase mode does not appear as a sharp in-gap resonance.
\end{abstract}

\maketitle
In semi-metals and narrow-band semiconductors, the electron-hole attraction can lead to excitonic insulator (EI) phase~\cite{des1965exciton,jerome1967excitonic,kunevs2015excitonic}, characterized by excitonic condensation that hybridizes conduction and valence bands. EI transition has similarities with the superconducting one but with an order parameter that is not charged. 
Textbook EIs involve the spontaneous breaking of U(1) symmetry associated with separate conservation of charge in valence/conduction orbitals and are realized in heterostructure/bilayer systems~\cite{Eisenstein2014,Li2017,Liu2017,Wang2019,Gupta2020,Nilsson2023}. 
 In bulk materials, the charge in orbital subspace is not conserved even prior to excitonic condensation and determining the broken symmetry is more nuanced~\cite{mazza2020nature}. 
Prominent candidates for bulk EI  are 1T-TiSe$_2$~\cite{cercellier2007,monney2010,kogar2017}, Sb nanoflakes~\cite{li2019}, Ta$_2$Pd$_3$Te$_5$~\cite{zhang2024,hossain2023,huang2024,yao2024} and  Ta$_2$NiSe$_5$ (TNS)~\cite{wakisaka2009excitonic}. 
 
TNS exhibits a resistivity anomaly at $T_{c}=325$K where the material undergoes a transition from semimetal to a small-gap $\Delta\sim0.15$eV semiconductor 
~\cite{di1986physical,lu2017zero,larkin2017giant,larkin2018infrared}. Simultaneously, a structural transition occurs from a high-temperature orthorhombic phase to a low-temperature monoclinic phase. The structural change is small with the monoclinic angle deviating from the high-symmetry configuration by a mere 0.7 degrees~\cite{sunshine1985structure}. 

Photoemission studies observed a flat valence band and argued that it arises from gap opening due to excitonic condensation~\cite{wakisaka2009excitonic,wakisaka2012photoemission,seki2014excitonic}. Additionally, short time scales and peculiar fluence dependencies observed in pump-probe time-resolved studies~\cite{Mor2017,mor2018,Okazaki2018,Bretscher2021,Bretscher2021b,Saha2021,geng2024} as well as the vanishing of thermopower at low temperatures~\cite{nakano2019exciton}  have been interpreted in terms of electronic degrees of freedom and argued to point to excitonic ground state.
A realistic theoretical investigation~\cite{mazza2020nature} 
identified excitonic order parameter $\phi$ with B$_{2g}$ pattern of hybridization between Ni and Ta-based orbitals, breaking discrete (mirror) symmetries of the orthorhombic phase. This study aligns with earlier considerations of excitonic phases obtained using simpler  models~\cite{kaneko2012electronic,kaneko2013orthorhombic}.

Another line of research highlights structural aspects~\cite{subedi2020orthorhombic,watson2020band,windgatter2021common} and argues the transition is due to freezing of an unstable B$_{2g}$ mode corresponding to a structural order parameter $X$. Some pump-probe angle-resolved photoemission spectroscopy (ARPES) studies~\cite{baldini2023spontaneous} argue against the presence of excitonic effects. Importantly, in equilibrium, $\phi$ and $X$ have the same symmetry. A nonvanishing $X$ activates real electronic hybridizations of the same kind as $\phi$.  This implies that the monoclinic distortion can explain the flatness of the bands or the vanishing of thermopower~\cite{nakano2019exciton}, challenging purely excitonic scenarios.

Electron-phonon coupling plays an  important role~\cite{larkin2017giant,larkin2018infrared,mor2018,ye2021lattice,golez2022,baldini2023spontaneous}.
Raman spectroscopy~\cite{Kim2020,kim2021direct,Volkov2021,volkov2021failed}  reveals a strongly asymmetric shape of the B$_{2g}$ mode, with some studies suggesting softening~\cite{Kim2020}, but all observing significant broadening of this phonon mode and hybridization with a strong electronic background as the temperature is lowered towards $T_c$.  
This broadening suggests that the coupling between electronic and lattice degrees of freedom above T$_c$ will manifest in some manner also below T$_c$.  
Several recent studies take a balanced perspective, emphasizing the relevance of both excitonic and structural aspects~\cite{yan2019strong,katsumi2023disentangling,tang2020non}. 
Very recent nonequilibrium Raman study~\cite{katsumi2023disentangling} found a nonequilibrium state without a gap but with monoclinic distortion. 

A crucial but so far poorly addressed question is that of the collective response~\cite{werdehausen2018coherent,Haque2024}. In this paper, we evaluate the excitonic order parameter susceptibility and Raman response in 
 a "minimal" six-band model~\cite{mazza2020nature}  that we extend by coupling to the relevant phonon mode with B$_{2g}$ symmetry.
We find several surprising features. The amplitude mode is not pinned to the lower edge of the single particle gap but has a larger energy. We find a strong phase mode at a finite frequency that stiffens further when the electron-phonon coupling is introduced. 
While our study focuses on Ta$_2$NiSe$_5$, its findings concerning collective behaviour in the presence of multiple bands, order parameter with symmetry lower than U(1), and coupling to phonon modes apply to other EI candidates, e.g. Ta$_2$Pd$_3$Te$_5$~\cite{zhang2024,hossain2023,huang2024,yao2024}. 

\begin{figure*}
\includegraphics[width=1.0\linewidth]{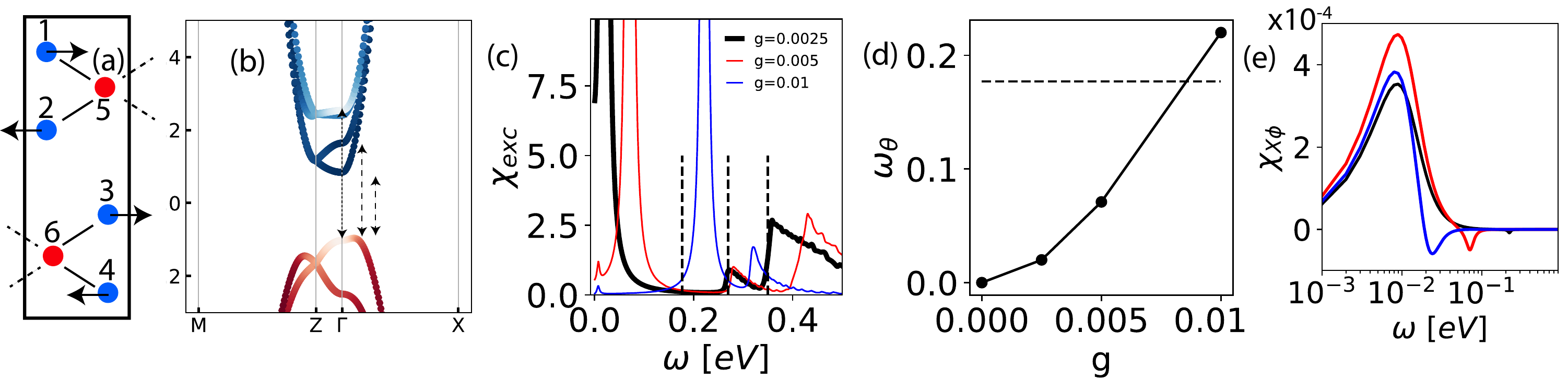}%
\caption{(a)~Unit cell with Ta~(blue) and Ni~(red) atoms. Arrows indicate B$_{2g}$ distortion.  The full~(dashed) lines mark the positive~(negative) value of the hybridization between Ta and Ni orbitals.(b)~Band-structure in the symmetry-broken state where blue~(red) shading encodes the weight of Ta~(Ni) orbitals.  (c)~Excitonic susceptibility $\chi_{\text{exc}}$ for a kick coupling to order parameter amplitude for different electron-phonon interactions $g$. The vertical dashed lines indicate energies marked in (b). (d)~The mass of the phase mode $\omega_{\theta}$ versus $g$. The dashed line marks the edge of the single-particle continuum.~(e) Cross-correlation function between excitonic and lattice order parameters for several $g$ from panel (c).  We have fixed the single-particle gap for all examples by adjusting the nearest neighbour electronic interactions; see SM for details. Parameters: $\beta=100$, damping $\eta=0.003$. }\label{fig1}
\end{figure*}

{\it Model and methods -- } Ta$_2$NiSe$_5$ is a layered compound. Within each layer in the 'ac' plane one encounters Ta-Ni-Ta chains that form a quasi-one dimensional structure.
We follow Ref.~\cite{mazza2020nature} by retaining the low-energy bands spanned by one Wannier orbital per each Ta and Ni atom and using the unit cell depicted in Fig.~\ref{fig1}(a). 
In the orthorhombic phase, the material has an inversion $\mathcal{I}$ and four reflection symmetries for planes parallel and perpendicular 
to the Ta-Ni-Ta chain~($\sigma_{\parallel,\perp}^{A/B}$) and the latter are broken as we enter the monoclinic phase~\cite{mazza2020nature,watson2020band}. 

The kinetic part of the Hamiltonian is represented in the tight-binding form
\begin{equation}
\hat H_{\text{kin}}=\sum_{\mathbf{R}\delta} \hat \Psi^{\dagger}_{\mathbf{R}+\delta\sigma} T(\delta) \hat \Psi_{\mathbf{R} \sigma}, 
\end{equation}
where the sum is taken over all unit cells $\mathbf{R}$ and neighbouring cells $\mathbf{\delta}$ as parametrized by the matrix of tight-binding hopping elements $T$, see SM for details. We have introduced the spinor $\hat \Psi_{\mathbf{R}\sigma}=\{\hat c_{1\sigma \mathbf{R}},\ldots,\hat c_{6\sigma \mathbf{R}}\}$ and $\hat c_{i\sigma \mathbf{R}}$ is the annihilation operator in unit cell $\mathbf{R}$, orbital $i$ and spin $\sigma$, see Fig.~\ref{fig1}(a). The electronic interaction  energy is given by 
\begin{eqnarray}
\hat H_{\text{int}}&=&U \sum_{\mathbf{R}i} \hat n_{i\uparrow \mathbf{R}}\hat n_{i\downarrow \mathbf{R}} +  V \sum_{\substack{i\in\{1,2\}\\ \mathbf{R}\sigma\sigma'}}  \hat n_{i\sigma \mathbf{R}+\delta_i} \hat n_{5\sigma' \mathbf{R}}\\&+&V \sum_{\substack{i\in\{3,4\}\\ \mathbf{R}\sigma\sigma'}}  \hat n_{i\sigma \mathbf{R}+\delta_i} \hat n_{6\sigma' \mathbf{R}}\nonumber,
\end{eqnarray}
where the first term represents the on-site Hubbard interaction for all atoms. The second and third terms are the nearest neighbour interactions, and the $\delta_i$ represents the sum over all nearest neighbours of both Ni atoms.

The electronic order parameters measure the amount of breaking of the four reflection symmetries, namely $\sigma_{\parallel,\perp}^{A/B}$. It is convenient to define operators corresponding to the electronic order parameter as $\hat \phi_\mathbf{R}=\{\hat \phi_{15\mathbf{R}},\hat \phi_{25\mathbf{R}},\hat \phi_{36\mathbf{R}},\hat \phi_{46\mathbf{R}}\},$ where $\hat \phi_{ij\mathbf{R}}= \sum_{\sigma}\hat c_{i\sigma \mathbf{R}}^{\dagger} \hat c_{j\sigma \mathbf{R}}+\hat c_{i\sigma \mathbf{R}\pm a}^{\dagger} \hat c_{j\sigma \mathbf{R}},$ where $a$ is a unit-cell displacement in the $x$ direction. We choose $+(-)$ for $\{ij\}=15,25~(36,46)$, respectively.
The electronic order parameter is then defined as the thermal expectation value of the operator $\phi_\mathbf{R}=\langle\hat{\phi}_\mathbf{R}\rangle$ and in equilibrium, all its components are real. In the following, we will focus on the experimentally relevant phase where expectation values have the same magnitude and their sign alternates as $\phi_\mathbf{R}=\phi_0 \{ 1,-1,-1,1\}$ breaking   $\sigma_{\parallel,\perp}^{A/B}$ symmetries consistent with the monoclinic distortion. 

Now, we will consider the coupling between electrons and lattice distortion corresponding to optical shearing mode with a frequency of 2 THz, see Fig.~\ref{fig1}(a), modelled as an Einstein phonon  $\hat H_{\text{ph}}=\frac{\omega_0}{2} \sum_\mathbf{R} (\hat X_\mathbf{R}^2+\hat \Pi_\mathbf{R}^2)$ where $\hat X_\mathbf{R}~(\hat \Pi_\mathbf{R})$ are the lattice distortion~(momentum) in the unit cell $\mathbf{R}$. In the monoclinic phase, the symmetry of the crystal structure is $C_{2h}$ with the order parameter transforming as the B$_{2g}$ irreducible representation and therefore the electronic order $\phi_{\bold{R}}$ and the lattice mode $X_{\bold{R}}$~(both in the $B_{2g}$ channel) are linearly coupled. A similar conclusion can be obtained by considering a Peirls-type of coupling, where larger~(smaller) nuclei distance leads to a smaller~(larger) absolute value of the hopping matrix element, see SM. For the distortion marked in Fig.~\ref{fig1}(a), it is easy to show the electron-lattice coupling reduces to
\begin{equation}\label{Eq:elph}
    \hat H_{\text{el-ph}}=g \sum_{\mathbf{R}} \hat X_\mathbf{R} (\hat \phi_{15\mathbf{R}}-\hat \phi_{25\mathbf{R}}-\hat \phi_{36\mathbf{R}}+\hat \phi_{46\mathbf{R}}),
\end{equation}
where we will treat the electron-phonon interaction $g$ as a free parameter. The system will lower its energy by forming an intertwined lattice and electronic order in the ground state~\cite{kaneko2013orthorhombic,sugimoto2016}. 

We solve the problem within the Hartree-Fock theory for the electron-electron and electron-lattice interaction within the symmetry-broken state and assume a homogenous solution~($\phi=\phi_\mathbf{R}$ and $X=X_\mathbf{R}$). We focus on the dynamical two-particle susceptibilities $\chi_{AB} (\omega)$ and consider three types of susceptibilities. The excitonic susceptibility $\chi_{\text{exc}}$ is given by the order parameter autocorrelation function  with $\hat A=\hat B=\hat \phi_{15}-\hat \phi_{25}-\hat \phi_{36}+\hat \phi_{46} + h.c. $ and measure the collective response of the excitonic order for a kick coupling to its amplitude direction (real part of order parameter; out of equilibrium $\phi$ becomes complex-valued). Furthermore, we can define the mixed lattice-exciton correlation function $\chi_{X\phi}$ with $\hat A=\hat \phi_{15}-\hat \phi_{25}-\hat \phi_{36}+\hat \phi_{46} + h.c.$ and $\hat B= \hat X$ which is a measure of correlations between the excitonic and lattice excitations. Finally, the electronic Raman response is determined using 
\begin{equation}
    \hat A=\hat B =  \sum_{\mathbf{\mathbf{R}}\sigma ij} t_{\mathbf{\mathbf{R}}i; \mathbf{0}j} (\mathbf{\mathbf{R}}_{ij}\cdot {\mathbf{e}}_1) (\mathbf{\mathbf{R}}_{ij}\cdot {\mathbf{e}}_2) e^{-i \mathbf{\mathbf{R}}\cdot \mathbf{k}}\hat c^\dagger_{i\sigma\mathbf{k}} \hat c_{j\sigma\mathbf{k}},
\end{equation}
where ${ \mathbf{e}}_i$ is the polarization of the Raman pulse and we will focus on ac channel, namely $\mathbf{e}_{1}=\{1,0\}$ and $\mathbf{e}_{2}=\{0,1\},$ see SM for details.
We take $k_B=\hbar=1$ and use eV as the unit of energy when not explicitly written otherwise. 

{\it Results -- }
\label{sec:summary}
We set the intra-band interaction $U=2.5$ and the inter-band interaction $V$=0.9 and consider the system at a low temperature with $\beta=1/T=100$ at which the system is in the excitonic phase, see also Ref.~\cite{mazza2020nature}. Fig.~\ref{fig1}(b) depicts the corresponding band structure and the excitonic hybridization opens up a gap of $160$ meV, which is close to the experimental values~\cite{lu2017zero,seki2014excitonic,larkin2017giant}. The orbital character of the bands is depicted with colours. One sees a strong admixture of Ta character in the valence bands close to the $\Gamma$ point. Remarkably, the excitonic hybridization does not strongly affect the lowest pair of conduction bands. Rather, it strongly affects the two higher conduction bands.

This observation has important consequences for the collective response. In Fig.~\ref{fig1}(c), we depict excitonic susceptibility $\chi_{\text{exc}}(\omega)$ for different electron-lattice couplings $g$, where we adjusted the nearest neighbours electronic interaction $V$ to fix the single-particle gap. For the weakest coupling $g=0.0025$, there are three dominant features: around $\omega\approx 350$ meV there is a strong shoulder whose energy matches the energy difference between highly hybridized bands marked in Fig.~\ref{fig1}(b) by a long arrow, a weaker shoulder at 270 meV and at very low energies $\omega\approx $20 meV a dominant peak. By comparison with the real-time evolution of the order parameter, we attribute the high energy peak to the amplitude mode and the low energy resonance to the phase mode~\cite{murakami2020collective}. However, due to the broken $U(1)$ symmetry, the two modes are coupled~\cite{murakami2017photoinduced}. The weak shoulder is not associated with the collective response. Its energy matches the transition between the top valence band and flat region on the second conducting band close to the $\Gamma$ point~(van Hove singularity).

Four interesting observations distinguish the realistic EI from the ideal U(1) symmetry-breaking one. Firstly, the phase mode that is massless in the ideal EI acquires a finite mass $\omega_{\theta}$ due to interchain hoppings that violate U(1) symmetry~\cite{mazza2020nature}. Its mass grows further with the electron-lattice coupling, see Fig.~\ref{fig1}(d) and can exceed the single-particle gap.  While it is difficult to determine the exact frequency of the massive phase mode due to the uncertainties in the microscopic parameters, we can estimate its lower boundary from the response in the purely electronic case which gives a value of order tens of the meV. Note that the origin of the finite mass is distinct from the case of superconductors where the phase mode acquires a large mass due to the Anderson-Higgs mechanism. Secondly, the amplitude mode is seen to oscillate with a frequency of 350 meV, which is larger than the single particle gap (in contrast to the standard case where one finds the amplitude mode pinned to the single particle gap~\cite{murakami2020collective}). This comes from the fact that the lowest conduction band is almost unaffected by the excitonic hybridization.  Thirdly, the "phase" and the "amplitude" mode are coupled. The response occurs in both modes following a kick coupling to the amplitude of the electronic order. Finally, for stronger electron-lattice coupled systems, e.g. $g\geq 0.005$, there is also a visible resonance at the phonon energy $\omega_0=8$ meV, see Fig.~\ref{fig1}(c). 

One can directly study the exciton-lattice coupling by considering the corresponding cross-correlation $\chi_{X\phi}$, see Fig.~\ref{fig1}(e). For all examples studied, there is a strong correlation between high-energy excitonic and low-energy lattice modes despite an order of magnitude different energy scales. Any distortion of the electronic order at high energies would strongly distort the lattice mode. We believe this is a natural explanation of recent optical pump-Thz probe experiments on Ta$_2$NiSe$_5$~\cite{michael2022,Haque2024} observing a strong THz amplification of reflectivity.

\begin{figure}
\includegraphics[width=1.0\linewidth]{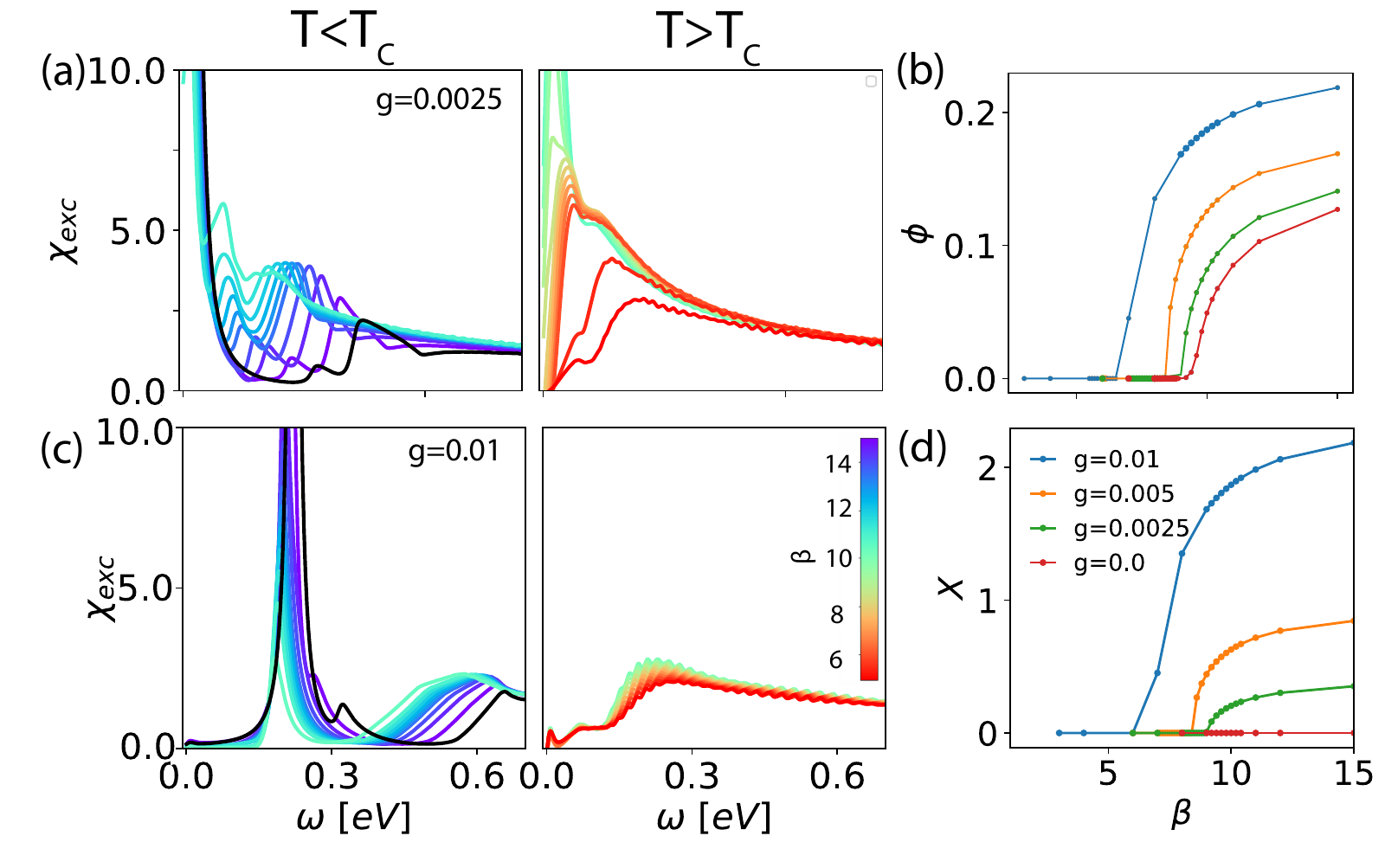}%
\caption{(a,c) The temperature dependence of the excitonic susceptibility $\chi_{\text{exc}}$ below (left) and above $T_c$ (right) for weak $g=0.0025$~(a) and strong $g=0.01$ electron-lattice coupling~(c). (b) and (d) shows the temperature dependence of the electronic order parameter $\phi$ and lattice distortion $X$, respectively. Damping parameter $\eta=0.01$.}
\label{fig2}
\end{figure}

Phenomenologically, we can understand the above response within the time-dependent Landau theory for lattice coupled with the real part of the electronic order parameter~\cite{golez2020}
$$ F[\phi]=  m|\dot{\phi}|^2+ a  |\phi|^2 + b |\phi|^4 +c(g) X |\phi| \cos(\theta) + d \cos{\theta} + F_{\text{ph}} , $$
where we have written the order parameter with amplitude and phase $\phi=|\phi| e^{-\I \theta}$ ($\theta$ may differ from 0 in time-dependent solution), $a,b,c,d,m$ are phenomenological parameters and $F_{\text{ph}}$ is the contribution of the free energy coming from the lattice. The coupling between the order parameters is described by $c(g)$ which monotonously increases with electron-lattice coupling strength $g$, see Ref.~\cite{golez2020} for explicit relations, and $d$ describes breaking of the $U(1)$ symmetry on the purely electronic level~\cite{mazza2020nature}. We see that the stronger the electron-lattice coupling, the larger the lattice distortion becomes, see also Fig.~\ref{fig2}(d), and the more the phase mode stiffens. Furthermore, the distortion of the excitonic amplitude mode acts as a force term on the lattice distortion, explaining the cross-correlation across large energy scales.

Further intuition is obtained by studying the excitonic susceptibility shown in Fig.~\ref{fig2}.  Panels (b) and (d) show the temperature dependence of $\phi$ and $X$, respectively. Due to the linear coupling, the two quantities vanish at the same $T_c$. The dimensionless lattice distortion $X$ is measured in $X_0=\sqrt{\hbar/(2 M \omega_0)}=0.07\AA$ (setting $M$ to reduced mass of Ta and Ni ion) and by comparison with experimental distortion of $X_{0,\mathrm{exp}}=0.04$  one infers electron-lattice coupling is small, e.g. $g=0.0025.$  For such $g$ the excitonic susceptibility above $T_c$ shows a clear softening of the excitonic response, see Fig.~\ref{fig2}(a), and for $T<T_C$ there is a gradual hardening of the amplitude and phase mode. On the contrary for large $g=0.01$ (panel (c)), the excitonic susceptibility does not show any softening above critical temperature $T>T_c$ and as we lower the temperature both the phase and the amplitude modes harden and the former gets substantially sharper. According to Ref.~\cite{volkov2021failed}, the experimental Raman data is consistent with the former scenario.

\begin{figure}
\includegraphics[width=1.0\linewidth]{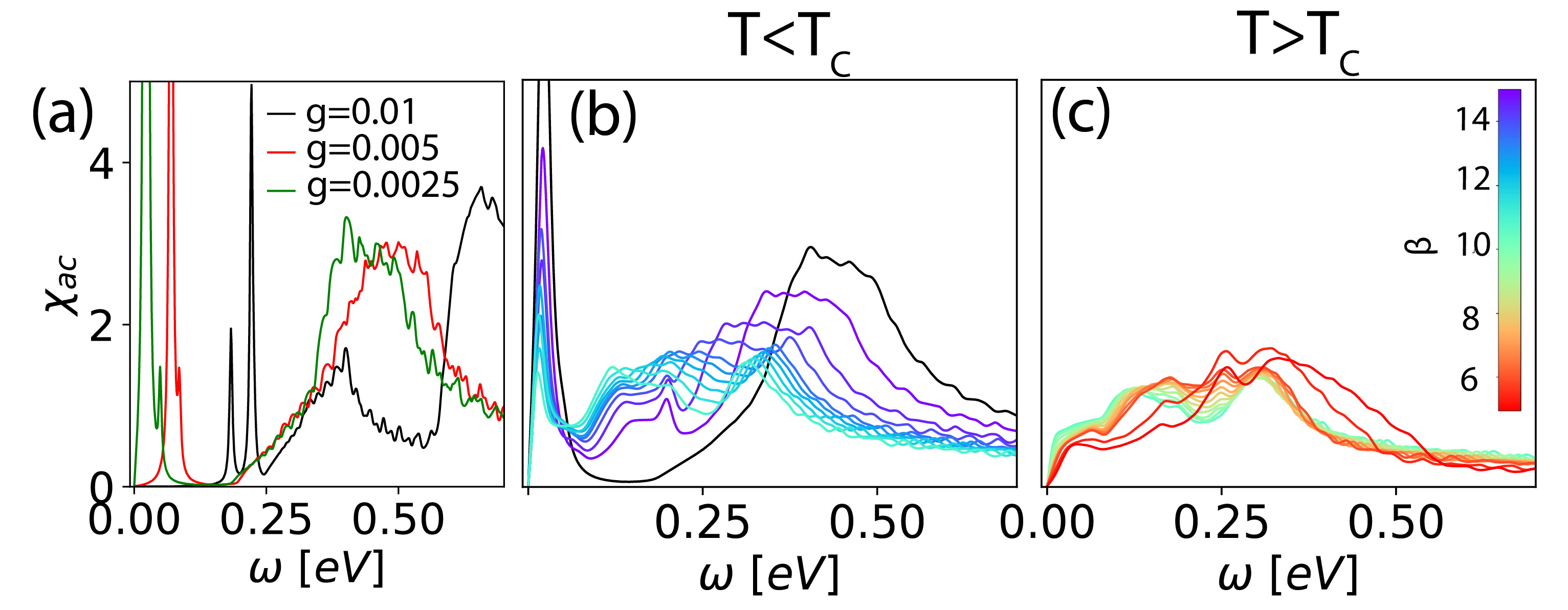}
\caption{(a)~Raman spectrum in the B$_{2g}$ channel~$\chi_\mathrm{ac}$ for several electron-lattice interactions $g$ in the low-temperature state at $\beta=100.$ Temperature dependence of the Raman spectrum for $g=0.0025$ below (b) and above $T_c$~(c).}
\label{fig3}
\end{figure}

We have evaluated Raman response within the Peierls approximation, see SM for details. The data shown in Fig.~\ref{fig3} exhibit similar behaviour as the order parameter susceptibility with strong and rather broad amplitude mode above the single-particle gap which gradually stiffens with decreasing temperatures and the phase mode, whose mass grows with increasing $g$. Like in the excitonic susceptibility one sees a softening of the electronic response as we approach $T_c$ from above and the hardening of the phase and amplitude mode as one approaches the ground state. 

The finding of amplitude mode above the single-particle gap is consistent with Raman experiments that do find a peak at about 400meV~\cite{Volkov2021}. Despite the amplitude mode being a rather broad resonance due to coupling with the continuum, we can still observe several coherent oscillations in the real-time signal~(not shown) and its strong sensitivity on the buildup of the order, see Fig.~\ref{fig2}. Our identification of the amplitude mode energy opens a promising perspective for probing it via nonlinear spectroscopy~\cite{tanabe2021} in analogy to superconductors~\cite{matsunaga2013,matsunaga2014,tsuji2015,blommel2024}. 

On the other hand, there is no sharp phase mode in the experimental data.
How can one reconcile the experimental data with our results?
The first observation is that we use the dissipationless Hartree-Fock approximation. Including the electron-lattice scattering should lead to a finite lifetime of the phase mode. As their coupling is non-linear, estimating its lifetime requires simulation with the inclusion of fluctuations effects, which goes beyond the scope of the present manuscript.
In Raman experiments, there indeed is a broad `excitonic continuum'~\cite{Volkov2021} with a gap of $\approx 20$ meV which could be a natural candidate for 
the overdamped phase mode. The second option is that the electron-lattice coupling is strong enough that the phase mode enters the single-particle continuum. Once again, the inclusion of scattering would strongly reduce its lifetime. This scenario is less probable as it is inconsistent with the experimental observation of electronic softening across the critical temperature and estimations of the lattice distortion. The last and most exotic option is that the phase mode acquires a larger mass due to long-range quadrupole-quadrupole interaction. Because the quadrupole interaction falls with the fifth power of the distance, we expect that such an effect is small, but should be nevertheless investigated in future.

In conclusion, we calculated collective and Raman responses in 
Ta$_2$NiSe$_5$. In contrast to the standard excitonic insulator response, we have identified the amplitude mode living well above the single-particle gap and matched the Raman spectroscopy data, opening the possibility of Higgs spectroscopy in this material. We find a massive phase mode with a mass that rapidly grows with the strength of the electron-lattice interaction. We argue that the response found in this work is generic for systems with coupled excitonic and lattice symmetry-broken states.  In the experiments, the phase mode is not observed (or at least is not a clear resonance) and we propose different scenarios for the discrepancy which remains an important problem for future research and should be resolved for the excitonic picture to be viable. 

\begin{acknowledgments}
{\it Acknowledgements}
We acknowledge support from No. P1-0044, No. J1-2455, No. J1-2458 and No. MN-0016-106 of the Slovenian Research Agency (ARIS). 
We acknowledge discussions with A. Georges, G. Mazza, and A. Subedi.
\end{acknowledgments}

\appendix
\newpage
\subsection{Details of the model}
In this work, we follow Ref.~\cite{mazza2020nature} which used density-functional theory input to construct a tight-binding Hamiltonian through the Wannierization technique as a minimal realistic model of a 2D layer of Ta$_2$NiSe$_5$. A two-dimensional problem with six bands gives the effective Hamiltonian, and they consider nearest neighbour hopping in the vertical and horizontal directions. The matrix elements of the kinetic energy $T_{\delta}$ are given in Table~\ref{tab:hopping}.

For later usage, it is convenient to 
write the interaction term in Eq.~\eqref{Eq.int} in a uniform form with full-density interaction as
$$ \hat H_\text{int}=\sum_{\textbf{k},\textbf{q}} V_{\textbf{q}}^{ij}  \hat n_{i\textbf{k}} \hat n_{j\textbf{k+q} },$$ where we have introduced the full density operator at momentum $\textbf{k}$ for orbital $i$ as $\hat n_{i\textbf{k}}=\sum_{\textbf{q}\sigma} \hat c_{i \sigma \textbf{k+q}}^{\dagger} \hat c_{i\sigma \textbf{q}}.$ The interaction vertex matrix is given by
\beq{
\label{Eq.int} 
V_{\textbf{k}}=\begin{pmatrix}
    U/2 & 0 & 0 &0 &V_1(\textbf{k}) & 0\\
    0 & U/2 & 0 &0 &V_1(\textbf{k}) & 0\\
    0 & 0 & U/2 &0 &0 & V_2(\textbf{k})\\
    0 & 0 & 0 &U/2 &0 & V_2(\textbf{k})\\
    V_1^*(\textbf{k}) & V_1^*(\textbf{k}) & 0 &0 &U/2 & 0\\
    0 & 0 & V_2^*(\textbf{k}) &V_2^*(\textbf{k}) &0 & U/2\\
\end{pmatrix},
}
with $V_1=V(1+\exp^{\I k_x})$ and $V_2=V(1+\exp^{-\I k_y})$.

\begin{table*}
\caption{\label{tab:hopping}Elements of the hopping matrix $T(\delta)$ obtained from Ref.~\cite{mazza2020nature} using first principle calculations and Wannierization.}
\begin{ruledtabular}
\begin{tabular}{c|lc}
 Cases& Hopping matrix elements t($\delta$)\\
\hline
Intra-chain Ta-Ta     & T$_{ii}$ (a$_{x}$,0)= T$_{ii}$ (-a$_{x}$,0)=-0.72 eV\\ 
                      &T$_{ii}$ (0,0)=1.35 eV, i=1,...,4\\
\hline
Intra-chain Ni-Ni& T$_{ii}$ (a$_{x}$,0)= T$_{ii}$ (-a$_{x}$,0)= 0.30 eV,\\
                & T$_{ii}$ (0,0)=-0.36eV, i=5,6\\
\hline
Intra-chain Ta-Ni& T$_{15}$(a$_{x}$,0)=T$_{25}$(a$_{x}$,0)=-T$_{25}$(0)=-T$_{15}$ (0)=-T$_{36}$(0)=-T$_{46}$(0)=T$_{36}$(-a$_{x}$,0),T$_{46}$ (-a$_{x}$,0)= -0.035 eV \\
\hline
Inter-chain Ta-Ni&T$_{35}$(a$_{x}$,0)=-T$_{35}$(-a$_{x}$,0),t$_{45}$(a$_{x}$,a$_{y}$)=-T$_{45}$(-a$_{x}$,a$_{y}              $)=-0.04 eV, \\
                & T$_{26}$(a$_{x}$,0)=-T$_{26}$(-a$_{x}$,0)=T$_{16}$(a$_{x}$,-a$_{y}$)=-T$_{16}$ (-a$_{x}$,-a$_{y}$)=T$_{45}$(-a$_{x}$,a$_{y}$)\\
\hline
Inter-chain Ta-Ta&T$_{23}$(0)=T$_{23}$(a$_{x}$,0)=T$_{41}$(-a$_{x}$,a$_{y}$)=T$_{41}$(0,a$_{y}$)=-0.02 eV \\
\hline
Inter-chain Ni-Ni& T$_{65}$(0)=T$_{65}$(a$_{x}$,0)=T$_{65}$(a$_{x}$,a$_{y}$)=T$_{65}$(0,a$_{y}$)=-0.03 eV \\
\end{tabular}
\end{ruledtabular}
\end{table*}

\subsection{Electron-lattice model}\label{sSec:El-lat}
Recent Raman studies~\cite{Kim2020,Volkov2021,volkov2021failed,ye2021lattice,kim2021direct} have unambiguously identified that the dominant lattice mode responsible for the orthorhombic to monoclinic transition is the $B_{2g}$ whose distortion pattern is marked in Fig.~\ref{fig1}(a). Here, we will model the electron-lattice coupling using the Peierls-like model, where the lattice distortion leads to a larger~(smaller) overlap between orbitals depending on the relative distance between atoms. Taking into account the phase arrangement of orbitals in real space, we can write the couplings between Ni 5 and Ta 1 from Fig.~\ref{fig1}(a) as
\bsplit{
	\hat H_{\text{el-ph}}&=g  \sum_{\textbf{R}\sigma} \hat X_{\textbf{R}} \big[-\hat c_{1\sigma \textbf{R}}^{\dagger} \hat c_{5\sigma \textbf{R}}+\hat c_{1\sigma \mathbf{R} +\mathbf{a}}^{\dagger} \hat c_{5\sigma \mathbf{R}}+\ldots \big],
}
where we have introduced the electron-phonon coupling $g$, the operator of the lattice distortion at site $\mathbf{R}$ in the B$_{2g}$ channel $\hat X_{\textbf{R}}.$ We can use similar arguments for other hoppings leading to Eq.~\ref{Eq:elph}.

\subsection{Method: Time-dependent Hartree-Fock}
We will employ the Hartree-Fock approximation for the ground-state 
and its time-dependent version for the evaluation of the susceptibility. Furthermore, we assume that the system is homogenous. The Hartree term is given by 
\beq{
  \Sigma^H_{ii}(t)=\begin{cases}
  V_{\textbf{k}=0}^{ii} n_{i}(t) \qquad \text{intraorbital} \\
  2V_{\textbf{k}=0}^{ij} n_{j}(t) \qquad \text{interorbital},
  \end{cases}
  \label{Eq:Hartree}
}
where we have assumed the local density $n_{i}=n_{i,\textbf{R}}$ and employed the Einstein notation.  The Fock term is given as 
\beq{
  \Sigma_{ij,\textbf{k}}^F(t)=-\sum_{\textbf{q}} V^{ij}_{\textbf{q}} \langle \hat \phi_{ij,\textbf{k-q}}\rangle(t)
  \label{Eq:Fock}
}

The electron-phonon interaction is described on a similar level as the electron-electron interaction with the decoupling 
\bsplit{
\hat{H}_{\text{el-ph}}=
-g \langle \hat X\rangle &\left[\hat \phi_{15}-\hat \phi_{25}-\hat \phi_{36}+\hat \phi_{46} + h.c. \right]\\
- g \hat X &\left[ \langle\hat\phi_{15}\rangle- \langle\hat\phi_{25}\rangle- \langle\hat\phi_{36}\rangle+\langle \hat\phi_{46}\rangle + h.c.\right],\nonumber
}
where the symbols with a hat marks the operator $\hat X$ and $X=\langle \hat X\rangle$ its expectation values.  Within the given approximation, the expectation value of the lattice follows the classical equation of motion given by
\bsplit{
	 \ddot  X=-\omega_0^2  X + g \omega_0  \left[ \phi_{15}- \phi_{25}- \phi_{36}+ \phi_{46} + h.c.\right]  
}
which for the excitonic order parameter in the $B_{2g}$ channel, namely $\phi=\phi_0\{1,-1,-1,1\}$  simplifies to
\bsplit{
	\ddot X=-\omega_0^2  X + 8 g \omega_0 \mathrm{Re}[\phi_0].
}

In equilibrium, we solve the Hartree-Fock equations by a standard self-consistency cycle. For real-time evolution, we propagate the density matrix as $\rho(t+dt)=e^{-\I \hat{H}^{}_{\text{eff}}[\rho] dt} \rho(t)e^{\I \hat{H}^{}_{\text{eff}}[\rho] dt},$ where $\hat{H}^{}_{\text{eff}}[\rho]$ presents an effective Hamiltonian including the kinetic, the Hartree-Fock and the electron-lattice contribution. The effective Hamiltonian $\hat{H}^{}_{\text{eff}}[\rho]$ depends on the single-particle density matrix and we perform a non-linear iteration at each timestep. The dominant computational complexity for the time propagation is the evaluation of the Fock self-energy, which includes an internal sum for each momentum. Here, we use the fast Fourier transform to evaluate the Fock term in the real space, leading to a linear complexity~(with logarithmic corrections) in the number of momentum points. The latter is crucial for the convergence of problems at long times. Finally, the equation of motion for the lattice degrees of freedom is solved using the multistep method. For the predictor we employ the 5th-order polynomial extrapolation and the corrector step is performed with the 5th-order Adams-Moulton method.

\subsection{Evaluation of two-particle susceptibilities}

One typically obtains the two-particle susceptibilities by solving the Bethe-Salpeter equation~\cite{onida2002}. However, as such  procedure is tedious in  multi-orbital and symmetry-broken systems~\cite{murakami2017photoinduced, murakami2020collective,kaneko2021}, we rather perform a time-dependent Hartree-Fock simulation where we add a time-dependent source field to the Hamiltonian~\cite{shao2016,golez2020}
\begin{equation}
  \hat H(t)= \hat H+\eta(t)  \hat A
 \label{tdhf}
\end{equation}
where we write the perturbation in terms of the operator $\hat A$ and its time dependence $\eta(t)$. In practice, we use a Gaussian pulse $\eta(t)=\eta_0 \exp^{-(t-t_0)^2/\sigma^2} $ with center at $t_0=4$ and the width $\sigma=1$. By restricting its height $\eta_0$ to be in the linear response regime, one can use the Kubo formalism and the susceptibility is given by the ratio of the Fourier transform for the observable $B$  and the source field $\chi_{BA}(\omega)=\langle \hat{B}\rangle(\omega)/\eta(\omega).$ 

\subsection{Comparison with Raman spectroscopy}
Tight binding Hamiltonian is $H_{\text{kin}}=\sum_{\textbf{k}} H(\textbf{k})$ with
\begin{equation}
\hat H(\textbf{k})=\sum_{\mathbf{R}\sigma ij} \hat c^{\dagger}_{i\sigma \mathbf{R}} \left[T_{ij}(\mathbf{R}) e^{-\I \mathbf{R}\cdot\mathbf{k}}  \right]\hat c_{j\sigma\mathbf{R}}, 
\end{equation}

The current (i.e. velocity) in the $\mathbf{e}$ direction is given by
\begin{equation}
    \mathbf{\hat{v}_k} \cdot \mathbf{e} = -i \sum_{\mathbf{R}\sigma ij} \hat c^\dagger_{i\sigma \mathbf{k}}\left[T_{ij}(\mathbf{R}) (\mathbf{R}_{ij}\cdot {\mathbf{e}}) e^{-i \mathbf{R}\cdot \mathbf{k}}\right] \hat c_{j\sigma \mathbf{k}},
\end{equation}
where $\mathbf{R}_{ij} = \mathbf{R}_i-\mathbf{R}_j$ and $\mathbf{R}_i = \mathbf{R} +\mathbf{r}_i$ with $\mathbf{r}_i$ the position of atomic orbital $i$ in the unit cell. 

Raman scattering operator for the incoming and outgoing polarization directions denoted by $\mathbf{e}_1$ and $\mathbf{e}_2$ is 
\begin{eqnarray}
    &{\mathbf{e}}_1 \cdot \hat{\cal S}{_\mathbf{k}} \cdot {\mathbf{e}}_2= \\ & \sum_{\mathbf{R}\sigma ij} \hat{c}^\dagger_{i\sigma \mathbf{k}} \left[t_{ij}(\textbf{R}) (\mathbf{R}_{ij}\cdot \mathbf{e}_1) (\mathbf{R}_{ij}\cdot \mathbf{e}_2) e^{-i \mathbf{R}\cdot \mathbf{k}}\right] \hat{c}_{j\sigma\mathbf{k}}.\nonumber
\end{eqnarray}
For aa scattering the normalized direction vectors ${\mathbf{e}}_{1,2}$ both point in the $\mathbf{a}$ direction, for ac scattering  $\mathbf{e}_1= \mathbf{a}/|\mathbf{a}|$ and $\mathbf{e}_2=\mathbf{c}/|\mathbf{c}|$.

\subsection{Bandstructure with increasing electron-lattice interaction}

\begin{figure}
\includegraphics[width=1.0\linewidth]{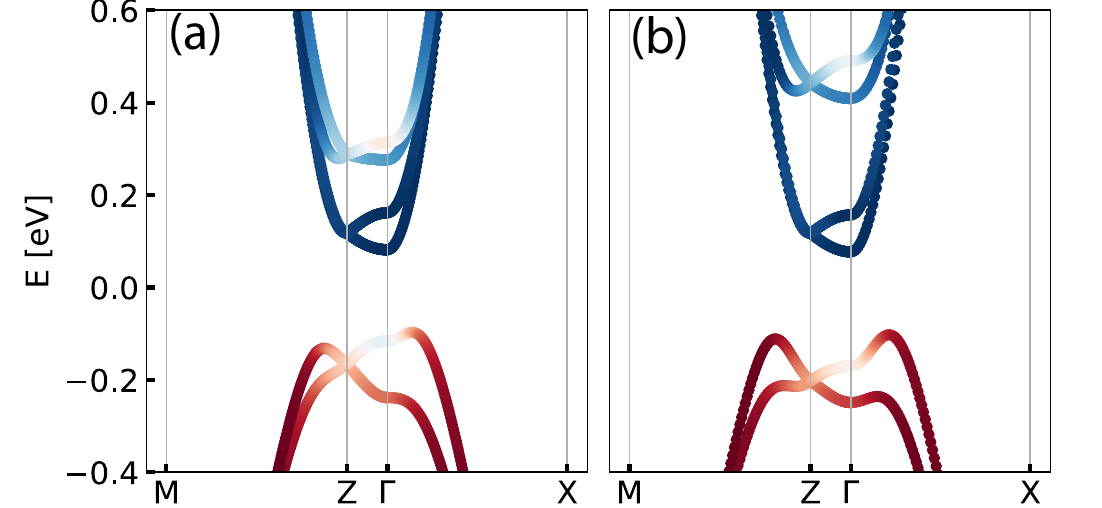}%
\caption{Band-structure in the symmetry-broken state where blue (red) shading encodes the weight of Ta (Ni) orbitals for the electron-lattice coupling $g=0.005$ and the nearest neighbour interaction $V=0.781$ eV~(a) and the electron-lattice coupling $g=0.01$ and the nearest neighbour interaction $V=0.7725$ eV~(b).}
\label{fig_sm}
\end{figure}

By increasing the electron-lattice interaction, naturally also the bandgap is modified. In order to make a fair comparison between different cases, we fix the size of the single-particle gap in the spectral function by adjusting the nearest neighbour interaction $V$ for increases in the electron-lattice interaction $g$, see Fig.~\ref{fig_sm}. The higher-lying mixed conducting band moves to higher energies consistent with the amplitude mode's evolution to higher energies with increasing electron-lattice coupling, see Fig.~\ref{fig1}(c).

\begin{thebibliography}{64}%
  \makeatletter
  \providecommand \@ifxundefined [1]{%
   \@ifx{#1\undefined}
  }%
  \providecommand \@ifnum [1]{%
   \ifnum #1\expandafter \@firstoftwo
   \else \expandafter \@secondoftwo
   \fi
  }%
  \providecommand \@ifx [1]{%
   \ifx #1\expandafter \@firstoftwo
   \else \expandafter \@secondoftwo
   \fi
  }%
  \providecommand \natexlab [1]{#1}%
  \providecommand \enquote  [1]{``#1''}%
  \providecommand \bibnamefont  [1]{#1}%
  \providecommand \bibfnamefont [1]{#1}%
  \providecommand \citenamefont [1]{#1}%
  \providecommand \href@noop [0]{\@secondoftwo}%
  \providecommand \href [0]{\begingroup \@sanitize@url \@href}%
  \providecommand \@href[1]{\@@startlink{#1}\@@href}%
  \providecommand \@@href[1]{\endgroup#1\@@endlink}%
  \providecommand \@sanitize@url [0]{\catcode `\\12\catcode `\$12\catcode
    `\&12\catcode `\#12\catcode `\^12\catcode `\_12\catcode `\%12\relax}%
  \providecommand \@@startlink[1]{}%
  \providecommand \@@endlink[0]{}%
  \providecommand \url  [0]{\begingroup\@sanitize@url \@url }%
  \providecommand \@url [1]{\endgroup\@href {#1}{\urlprefix }}%
  \providecommand \urlprefix  [0]{URL }%
  \providecommand \Eprint [0]{\href }%
  \providecommand \doibase [0]{https://doi.org/}%
  \providecommand \selectlanguage [0]{\@gobble}%
  \providecommand \bibinfo  [0]{\@secondoftwo}%
  \providecommand \bibfield  [0]{\@secondoftwo}%
  \providecommand \translation [1]{[#1]}%
  \providecommand \BibitemOpen [0]{}%
  \providecommand \bibitemStop [0]{}%
  \providecommand \bibitemNoStop [0]{.\EOS\space}%
  \providecommand \EOS [0]{\spacefactor3000\relax}%
  \providecommand \BibitemShut  [1]{\csname bibitem#1\endcsname}%
  \let\auto@bib@innerbib\@empty
  \bibitem [{\citenamefont {Des~Cloizeaux}(1965)}]{des1965exciton}%
    \BibitemOpen
    \bibfield  {author} {\bibinfo {author} {\bibfnamefont {J.}~\bibnamefont
    {Des~Cloizeaux}},\ }\href@noop {} {\bibfield  {journal} {\bibinfo  {journal}
    {Journal of Physics and Chemistry of Solids}\ }\textbf {\bibinfo {volume}
    {26}},\ \bibinfo {pages} {259} (\bibinfo {year} {1965})}\BibitemShut
    {NoStop}%
  \bibitem [{\citenamefont {J{\'e}rome}\ \emph {et~al.}(1967)\citenamefont
    {J{\'e}rome}, \citenamefont {Rice},\ and\ \citenamefont
    {Kohn}}]{jerome1967excitonic}%
    \BibitemOpen
    \bibfield  {author} {\bibinfo {author} {\bibfnamefont {D.}~\bibnamefont
    {J{\'e}rome}}, \bibinfo {author} {\bibfnamefont {T.}~\bibnamefont {Rice}},\
    and\ \bibinfo {author} {\bibfnamefont {W.}~\bibnamefont {Kohn}},\ }\href@noop
    {} {\bibfield  {journal} {\bibinfo  {journal} {Physical Review}\ }\textbf
    {\bibinfo {volume} {158}},\ \bibinfo {pages} {462} (\bibinfo {year}
    {1967})}\BibitemShut {NoStop}%
  \bibitem [{\citenamefont {Kune{\v{s}}}(2015)}]{kunevs2015excitonic}%
    \BibitemOpen
    \bibfield  {author} {\bibinfo {author} {\bibfnamefont {J.}~\bibnamefont
    {Kune{\v{s}}}},\ }\href@noop {} {\bibfield  {journal} {\bibinfo  {journal}
    {Journal of Physics: Condensed Matter}\ }\textbf {\bibinfo {volume} {27}},\
    \bibinfo {pages} {333201} (\bibinfo {year} {2015})}\BibitemShut {NoStop}%
  \bibitem [{\citenamefont {Eisenstein}(2014)}]{Eisenstein2014}%
    \BibitemOpen
    \bibfield  {author} {\bibinfo {author} {\bibfnamefont {J.}~\bibnamefont
    {Eisenstein}},\ }\href
    {https://doi.org/10.1146/annurev-conmatphys-031113-133832} {\bibfield
    {journal} {\bibinfo  {journal} {Annual Review of Condensed Matter Physics}\
    }\textbf {\bibinfo {volume} {5}},\ \bibinfo {pages} {159} (\bibinfo {year}
    {2014})}\BibitemShut {NoStop}%
  \bibitem [{\citenamefont {Li}\ \emph {et~al.}(2017)\citenamefont {Li},
    \citenamefont {Taniguchi}, \citenamefont {Watanabe}, \citenamefont {Hone},\
    and\ \citenamefont {Dean}}]{Li2017}%
    \BibitemOpen
    \bibfield  {author} {\bibinfo {author} {\bibfnamefont {J.~I.~A.}\
    \bibnamefont {Li}}, \bibinfo {author} {\bibfnamefont {T.}~\bibnamefont
    {Taniguchi}}, \bibinfo {author} {\bibfnamefont {K.}~\bibnamefont {Watanabe}},
    \bibinfo {author} {\bibfnamefont {J.}~\bibnamefont {Hone}},\ and\ \bibinfo
    {author} {\bibfnamefont {C.~R.}\ \bibnamefont {Dean}},\ }\href
    {https://doi.org/10.1038/nphys4140} {\bibfield  {journal} {\bibinfo
    {journal} {Nature Physics}\ }\textbf {\bibinfo {volume} {13}},\ \bibinfo
    {pages} {751} (\bibinfo {year} {2017})}\BibitemShut {NoStop}%
  \bibitem [{\citenamefont {Liu}\ \emph {et~al.}(2017)\citenamefont {Liu},
    \citenamefont {Watanabe}, \citenamefont {Taniguchi}, \citenamefont
    {Halperin},\ and\ \citenamefont {Kim}}]{Liu2017}%
    \BibitemOpen
    \bibfield  {author} {\bibinfo {author} {\bibfnamefont {X.}~\bibnamefont
    {Liu}}, \bibinfo {author} {\bibfnamefont {K.}~\bibnamefont {Watanabe}},
    \bibinfo {author} {\bibfnamefont {T.}~\bibnamefont {Taniguchi}}, \bibinfo
    {author} {\bibfnamefont {B.~I.}\ \bibnamefont {Halperin}},\ and\ \bibinfo
    {author} {\bibfnamefont {P.}~\bibnamefont {Kim}},\ }\href
    {https://doi.org/10.1038/nphys4116} {\bibfield  {journal} {\bibinfo
    {journal} {Nature Physics}\ }\textbf {\bibinfo {volume} {13}},\ \bibinfo
    {pages} {746} (\bibinfo {year} {2017})}\BibitemShut {NoStop}%
  \bibitem [{\citenamefont {Wang}\ \emph {et~al.}(2019)\citenamefont {Wang},
    \citenamefont {Rhodes}, \citenamefont {Watanabe}, \citenamefont {Taniguchi},
    \citenamefont {Hone}, \citenamefont {Shan},\ and\ \citenamefont
    {Mak}}]{Wang2019}%
    \BibitemOpen
    \bibfield  {author} {\bibinfo {author} {\bibfnamefont {Z.}~\bibnamefont
    {Wang}}, \bibinfo {author} {\bibfnamefont {D.~A.}\ \bibnamefont {Rhodes}},
    \bibinfo {author} {\bibfnamefont {K.}~\bibnamefont {Watanabe}}, \bibinfo
    {author} {\bibfnamefont {T.}~\bibnamefont {Taniguchi}}, \bibinfo {author}
    {\bibfnamefont {J.~C.}\ \bibnamefont {Hone}}, \bibinfo {author}
    {\bibfnamefont {J.}~\bibnamefont {Shan}},\ and\ \bibinfo {author}
    {\bibfnamefont {K.~F.}\ \bibnamefont {Mak}},\ }\href
    {https://doi.org/10.1038/s41586-019-1591-7} {\bibfield  {journal} {\bibinfo
    {journal} {Nature}\ }\textbf {\bibinfo {volume} {574}},\ \bibinfo {pages}
    {76} (\bibinfo {year} {2019})}\BibitemShut {NoStop}%
  \bibitem [{\citenamefont {Gupta}\ \emph {et~al.}(2020)\citenamefont {Gupta},
    \citenamefont {Kutana},\ and\ \citenamefont {Yakobson}}]{Gupta2020}%
    \BibitemOpen
    \bibfield  {author} {\bibinfo {author} {\bibfnamefont {S.}~\bibnamefont
    {Gupta}}, \bibinfo {author} {\bibfnamefont {A.}~\bibnamefont {Kutana}},\ and\
    \bibinfo {author} {\bibfnamefont {B.~I.}\ \bibnamefont {Yakobson}},\
    }\bibfield  {journal} {\bibinfo  {journal} {Nature Communications}\ }\textbf
    {\bibinfo {volume} {11}},\ \href {https://doi.org/10.1038/s41467-020-16737-0}
    {10.1038/s41467-020-16737-0} (\bibinfo {year} {2020})\BibitemShut {NoStop}%
  \bibitem [{\citenamefont {Nilsson}\ \emph {et~al.}(2023)\citenamefont
    {Nilsson}, \citenamefont {Kuisma}, \citenamefont {Pakdel},\ and\
    \citenamefont {Thygesen}}]{Nilsson2023}%
    \BibitemOpen
    \bibfield  {author} {\bibinfo {author} {\bibfnamefont {F.}~\bibnamefont
    {Nilsson}}, \bibinfo {author} {\bibfnamefont {M.}~\bibnamefont {Kuisma}},
    \bibinfo {author} {\bibfnamefont {S.}~\bibnamefont {Pakdel}},\ and\ \bibinfo
    {author} {\bibfnamefont {K.~S.}\ \bibnamefont {Thygesen}},\ }\href
    {https://doi.org/10.1021/acs.jpclett.3c00090} {\bibfield  {journal} {\bibinfo
     {journal} {The Journal of Physical Chemistry Letters}\ }\textbf {\bibinfo
    {volume} {14}},\ \bibinfo {pages} {2277} (\bibinfo {year}
    {2023})}\BibitemShut {NoStop}%
  \bibitem [{\citenamefont {Mazza}\ \emph {et~al.}(2020)\citenamefont {Mazza},
    \citenamefont {R{\"o}sner}, \citenamefont {Windg{\"a}tter}, \citenamefont
    {Latini}, \citenamefont {H{\"u}bener}, \citenamefont {Millis}, \citenamefont
    {Rubio},\ and\ \citenamefont {Georges}}]{mazza2020nature}%
    \BibitemOpen
    \bibfield  {author} {\bibinfo {author} {\bibfnamefont {G.}~\bibnamefont
    {Mazza}}, \bibinfo {author} {\bibfnamefont {M.}~\bibnamefont {R{\"o}sner}},
    \bibinfo {author} {\bibfnamefont {L.}~\bibnamefont {Windg{\"a}tter}},
    \bibinfo {author} {\bibfnamefont {S.}~\bibnamefont {Latini}}, \bibinfo
    {author} {\bibfnamefont {H.}~\bibnamefont {H{\"u}bener}}, \bibinfo {author}
    {\bibfnamefont {A.~J.}\ \bibnamefont {Millis}}, \bibinfo {author}
    {\bibfnamefont {A.}~\bibnamefont {Rubio}},\ and\ \bibinfo {author}
    {\bibfnamefont {A.}~\bibnamefont {Georges}},\ }\href@noop {} {\bibfield
    {journal} {\bibinfo  {journal} {Physical Review Letters}\ }\textbf {\bibinfo
    {volume} {124}},\ \bibinfo {pages} {197601} (\bibinfo {year}
    {2020})}\BibitemShut {NoStop}%
  \bibitem [{\citenamefont {Cercellier}\ \emph {et~al.}(2007)\citenamefont
    {Cercellier}, \citenamefont {Monney}, \citenamefont {Clerc}, \citenamefont
    {Battaglia}, \citenamefont {Despont}, \citenamefont {Garnier}, \citenamefont
    {Beck}, \citenamefont {Aebi}, \citenamefont {Patthey}, \citenamefont
    {Berger},\ and\ \citenamefont {Forr\'o}}]{cercellier2007}%
    \BibitemOpen
    \bibfield  {author} {\bibinfo {author} {\bibfnamefont {H.}~\bibnamefont
    {Cercellier}}, \bibinfo {author} {\bibfnamefont {C.}~\bibnamefont {Monney}},
    \bibinfo {author} {\bibfnamefont {F.}~\bibnamefont {Clerc}}, \bibinfo
    {author} {\bibfnamefont {C.}~\bibnamefont {Battaglia}}, \bibinfo {author}
    {\bibfnamefont {L.}~\bibnamefont {Despont}}, \bibinfo {author} {\bibfnamefont
    {M.~G.}\ \bibnamefont {Garnier}}, \bibinfo {author} {\bibfnamefont
    {H.}~\bibnamefont {Beck}}, \bibinfo {author} {\bibfnamefont {P.}~\bibnamefont
    {Aebi}}, \bibinfo {author} {\bibfnamefont {L.}~\bibnamefont {Patthey}},
    \bibinfo {author} {\bibfnamefont {H.}~\bibnamefont {Berger}},\ and\ \bibinfo
    {author} {\bibfnamefont {L.}~\bibnamefont {Forr\'o}},\ }\href
    {https://doi.org/10.1103/PhysRevLett.99.146403} {\bibfield  {journal}
    {\bibinfo  {journal} {Phys. Rev. Lett.}\ }\textbf {\bibinfo {volume} {99}},\
    \bibinfo {pages} {146403} (\bibinfo {year} {2007})}\BibitemShut {NoStop}%
  \bibitem [{\citenamefont {Monney}\ \emph {et~al.}(2010)\citenamefont {Monney},
    \citenamefont {Schwier}, \citenamefont {Garnier}, \citenamefont {Mariotti},
    \citenamefont {Didiot}, \citenamefont {Beck}, \citenamefont {Aebi},
    \citenamefont {Cercellier}, \citenamefont {Marcus}, \citenamefont
    {Battaglia}, \citenamefont {Berger},\ and\ \citenamefont
    {Titov}}]{monney2010}%
    \BibitemOpen
    \bibfield  {author} {\bibinfo {author} {\bibfnamefont {C.}~\bibnamefont
    {Monney}}, \bibinfo {author} {\bibfnamefont {E.~F.}\ \bibnamefont {Schwier}},
    \bibinfo {author} {\bibfnamefont {M.~G.}\ \bibnamefont {Garnier}}, \bibinfo
    {author} {\bibfnamefont {N.}~\bibnamefont {Mariotti}}, \bibinfo {author}
    {\bibfnamefont {C.}~\bibnamefont {Didiot}}, \bibinfo {author} {\bibfnamefont
    {H.}~\bibnamefont {Beck}}, \bibinfo {author} {\bibfnamefont {P.}~\bibnamefont
    {Aebi}}, \bibinfo {author} {\bibfnamefont {H.}~\bibnamefont {Cercellier}},
    \bibinfo {author} {\bibfnamefont {J.}~\bibnamefont {Marcus}}, \bibinfo
    {author} {\bibfnamefont {C.}~\bibnamefont {Battaglia}}, \bibinfo {author}
    {\bibfnamefont {H.}~\bibnamefont {Berger}},\ and\ \bibinfo {author}
    {\bibfnamefont {A.~N.}\ \bibnamefont {Titov}},\ }\href
    {https://doi.org/10.1103/PhysRevB.81.155104} {\bibfield  {journal} {\bibinfo
    {journal} {Phys. Rev. B}\ }\textbf {\bibinfo {volume} {81}},\ \bibinfo
    {pages} {155104} (\bibinfo {year} {2010})}\BibitemShut {NoStop}%
  \bibitem [{\citenamefont {Kogar}\ \emph {et~al.}(2017)\citenamefont {Kogar},
    \citenamefont {Rak}, \citenamefont {Vig}, \citenamefont {Husain},
    \citenamefont {Flicker}, \citenamefont {Joe}, \citenamefont {Venema},
    \citenamefont {MacDougall}, \citenamefont {Chiang}, \citenamefont {Fradkin}
    \emph {et~al.}}]{kogar2017}%
    \BibitemOpen
    \bibfield  {author} {\bibinfo {author} {\bibfnamefont {A.}~\bibnamefont
    {Kogar}}, \bibinfo {author} {\bibfnamefont {M.~S.}\ \bibnamefont {Rak}},
    \bibinfo {author} {\bibfnamefont {S.}~\bibnamefont {Vig}}, \bibinfo {author}
    {\bibfnamefont {A.~A.}\ \bibnamefont {Husain}}, \bibinfo {author}
    {\bibfnamefont {F.}~\bibnamefont {Flicker}}, \bibinfo {author} {\bibfnamefont
    {Y.~I.}\ \bibnamefont {Joe}}, \bibinfo {author} {\bibfnamefont
    {L.}~\bibnamefont {Venema}}, \bibinfo {author} {\bibfnamefont {G.~J.}\
    \bibnamefont {MacDougall}}, \bibinfo {author} {\bibfnamefont {T.~C.}\
    \bibnamefont {Chiang}}, \bibinfo {author} {\bibfnamefont {E.}~\bibnamefont
    {Fradkin}}, \emph {et~al.},\ }\href@noop {} {\bibfield  {journal} {\bibinfo
    {journal} {Science}\ }\textbf {\bibinfo {volume} {358}},\ \bibinfo {pages}
    {1314} (\bibinfo {year} {2017})}\BibitemShut {NoStop}%
  \bibitem [{\citenamefont {Li}\ \emph {et~al.}(2019)\citenamefont {Li},
    \citenamefont {Nadeem}, \citenamefont {Yue}, \citenamefont {Cortie},
    \citenamefont {Fuhrer},\ and\ \citenamefont {Wang}}]{li2019}%
    \BibitemOpen
    \bibfield  {author} {\bibinfo {author} {\bibfnamefont {Z.}~\bibnamefont
    {Li}}, \bibinfo {author} {\bibfnamefont {M.}~\bibnamefont {Nadeem}}, \bibinfo
    {author} {\bibfnamefont {Z.}~\bibnamefont {Yue}}, \bibinfo {author}
    {\bibfnamefont {D.}~\bibnamefont {Cortie}}, \bibinfo {author} {\bibfnamefont
    {M.}~\bibnamefont {Fuhrer}},\ and\ \bibinfo {author} {\bibfnamefont
    {X.}~\bibnamefont {Wang}},\ }\href@noop {} {\bibfield  {journal} {\bibinfo
    {journal} {Nano letters}\ }\textbf {\bibinfo {volume} {19}},\ \bibinfo
    {pages} {4960} (\bibinfo {year} {2019})}\BibitemShut {NoStop}%
  \bibitem [{\citenamefont {Zhang}\ \emph {et~al.}(2024)\citenamefont {Zhang},
    \citenamefont {Dong}, \citenamefont {Yan}, \citenamefont {Jiang},
    \citenamefont {Yang}, \citenamefont {Li}, \citenamefont {Guo}, \citenamefont
    {Huang}, \citenamefont {Li}, \citenamefont {Li} \emph {et~al.}}]{zhang2024}%
    \BibitemOpen
    \bibfield  {author} {\bibinfo {author} {\bibfnamefont {P.}~\bibnamefont
    {Zhang}}, \bibinfo {author} {\bibfnamefont {Y.}~\bibnamefont {Dong}},
    \bibinfo {author} {\bibfnamefont {D.}~\bibnamefont {Yan}}, \bibinfo {author}
    {\bibfnamefont {B.}~\bibnamefont {Jiang}}, \bibinfo {author} {\bibfnamefont
    {T.}~\bibnamefont {Yang}}, \bibinfo {author} {\bibfnamefont {J.}~\bibnamefont
    {Li}}, \bibinfo {author} {\bibfnamefont {Z.}~\bibnamefont {Guo}}, \bibinfo
    {author} {\bibfnamefont {Y.}~\bibnamefont {Huang}}, \bibinfo {author}
    {\bibfnamefont {Q.}~\bibnamefont {Li}}, \bibinfo {author} {\bibfnamefont
    {Y.}~\bibnamefont {Li}}, \emph {et~al.},\ }\href@noop {} {\bibfield
    {journal} {\bibinfo  {journal} {Physical Review X}\ }\textbf {\bibinfo
    {volume} {14}},\ \bibinfo {pages} {011047} (\bibinfo {year}
    {2024})}\BibitemShut {NoStop}%
  \bibitem [{\citenamefont {Hossain}\ \emph {et~al.}(2023)\citenamefont
    {Hossain}, \citenamefont {Cochran}, \citenamefont {Jiang}, \citenamefont
    {Zhang}, \citenamefont {Wu}, \citenamefont {Liu}, \citenamefont {Zheng},
    \citenamefont {Kim}, \citenamefont {Cheng}, \citenamefont {Zhang} \emph
    {et~al.}}]{hossain2023}%
    \BibitemOpen
    \bibfield  {author} {\bibinfo {author} {\bibfnamefont {M.~S.}\ \bibnamefont
    {Hossain}}, \bibinfo {author} {\bibfnamefont {T.~A.}\ \bibnamefont
    {Cochran}}, \bibinfo {author} {\bibfnamefont {Y.-X.}\ \bibnamefont {Jiang}},
    \bibinfo {author} {\bibfnamefont {S.}~\bibnamefont {Zhang}}, \bibinfo
    {author} {\bibfnamefont {H.}~\bibnamefont {Wu}}, \bibinfo {author}
    {\bibfnamefont {X.}~\bibnamefont {Liu}}, \bibinfo {author} {\bibfnamefont
    {X.}~\bibnamefont {Zheng}}, \bibinfo {author} {\bibfnamefont
    {B.}~\bibnamefont {Kim}}, \bibinfo {author} {\bibfnamefont {G.}~\bibnamefont
    {Cheng}}, \bibinfo {author} {\bibfnamefont {Q.}~\bibnamefont {Zhang}}, \emph
    {et~al.},\ }\href@noop {} {\bibfield  {journal} {\bibinfo  {journal} {arXiv
    preprint arXiv:2312.15862}\ } (\bibinfo {year} {2023})}\BibitemShut {NoStop}%
  \bibitem [{\citenamefont {Huang}\ \emph {et~al.}(2024)\citenamefont {Huang},
    \citenamefont {Jiang}, \citenamefont {Yao}, \citenamefont {Yan},
    \citenamefont {Lei}, \citenamefont {Gao}, \citenamefont {Guo}, \citenamefont
    {Jin}, \citenamefont {Li}, \citenamefont {Yuan} \emph {et~al.}}]{huang2024}%
    \BibitemOpen
    \bibfield  {author} {\bibinfo {author} {\bibfnamefont {J.}~\bibnamefont
    {Huang}}, \bibinfo {author} {\bibfnamefont {B.}~\bibnamefont {Jiang}},
    \bibinfo {author} {\bibfnamefont {J.}~\bibnamefont {Yao}}, \bibinfo {author}
    {\bibfnamefont {D.}~\bibnamefont {Yan}}, \bibinfo {author} {\bibfnamefont
    {X.}~\bibnamefont {Lei}}, \bibinfo {author} {\bibfnamefont {J.}~\bibnamefont
    {Gao}}, \bibinfo {author} {\bibfnamefont {Z.}~\bibnamefont {Guo}}, \bibinfo
    {author} {\bibfnamefont {F.}~\bibnamefont {Jin}}, \bibinfo {author}
    {\bibfnamefont {Y.}~\bibnamefont {Li}}, \bibinfo {author} {\bibfnamefont
    {Z.}~\bibnamefont {Yuan}}, \emph {et~al.},\ }\href@noop {} {\bibfield
    {journal} {\bibinfo  {journal} {Physical Review X}\ }\textbf {\bibinfo
    {volume} {14}},\ \bibinfo {pages} {011046} (\bibinfo {year}
    {2024})}\BibitemShut {NoStop}%
  \bibitem [{\citenamefont {Yao}\ \emph {et~al.}(2024)\citenamefont {Yao},
    \citenamefont {Sheng}, \citenamefont {Zhang}, \citenamefont {Wu},
    \citenamefont {Weng}, \citenamefont {Dai}, \citenamefont {Fang},\ and\
    \citenamefont {Wang}}]{yao2024}%
    \BibitemOpen
    \bibfield  {author} {\bibinfo {author} {\bibfnamefont {J.}~\bibnamefont
    {Yao}}, \bibinfo {author} {\bibfnamefont {H.}~\bibnamefont {Sheng}}, \bibinfo
    {author} {\bibfnamefont {R.}~\bibnamefont {Zhang}}, \bibinfo {author}
    {\bibfnamefont {Q.}~\bibnamefont {Wu}}, \bibinfo {author} {\bibfnamefont
    {H.}~\bibnamefont {Weng}}, \bibinfo {author} {\bibfnamefont {X.}~\bibnamefont
    {Dai}}, \bibinfo {author} {\bibfnamefont {Z.}~\bibnamefont {Fang}},\ and\
    \bibinfo {author} {\bibfnamefont {Z.}~\bibnamefont {Wang}},\ }\href@noop {}
    {\bibfield  {journal} {\bibinfo  {journal} {arXiv preprint arXiv:2401.01222}\
    } (\bibinfo {year} {2024})}\BibitemShut {NoStop}%
  \bibitem [{\citenamefont {Wakisaka}\ \emph {et~al.}(2009)\citenamefont
    {Wakisaka}, \citenamefont {Sudayama}, \citenamefont {Takubo}, \citenamefont
    {Mizokawa}, \citenamefont {Arita}, \citenamefont {Namatame}, \citenamefont
    {Taniguchi}, \citenamefont {Katayama}, \citenamefont {Nohara},\ and\
    \citenamefont {Takagi}}]{wakisaka2009excitonic}%
    \BibitemOpen
    \bibfield  {author} {\bibinfo {author} {\bibfnamefont {Y.}~\bibnamefont
    {Wakisaka}}, \bibinfo {author} {\bibfnamefont {T.}~\bibnamefont {Sudayama}},
    \bibinfo {author} {\bibfnamefont {K.}~\bibnamefont {Takubo}}, \bibinfo
    {author} {\bibfnamefont {T.}~\bibnamefont {Mizokawa}}, \bibinfo {author}
    {\bibfnamefont {M.}~\bibnamefont {Arita}}, \bibinfo {author} {\bibfnamefont
    {H.}~\bibnamefont {Namatame}}, \bibinfo {author} {\bibfnamefont
    {M.}~\bibnamefont {Taniguchi}}, \bibinfo {author} {\bibfnamefont
    {N.}~\bibnamefont {Katayama}}, \bibinfo {author} {\bibfnamefont
    {M.}~\bibnamefont {Nohara}},\ and\ \bibinfo {author} {\bibfnamefont
    {H.}~\bibnamefont {Takagi}},\ }\href@noop {} {\bibfield  {journal} {\bibinfo
    {journal} {Physical review letters}\ }\textbf {\bibinfo {volume} {103}},\
    \bibinfo {pages} {026402} (\bibinfo {year} {2009})}\BibitemShut {NoStop}%
  \bibitem [{\citenamefont {Di~Salvo}\ \emph {et~al.}(1986)\citenamefont
    {Di~Salvo}, \citenamefont {Chen}, \citenamefont {Fleming}, \citenamefont
    {Waszczak}, \citenamefont {Dunn}, \citenamefont {Sunshine},\ and\
    \citenamefont {Ibers}}]{di1986physical}%
    \BibitemOpen
    \bibfield  {author} {\bibinfo {author} {\bibfnamefont {F.}~\bibnamefont
    {Di~Salvo}}, \bibinfo {author} {\bibfnamefont {C.}~\bibnamefont {Chen}},
    \bibinfo {author} {\bibfnamefont {R.}~\bibnamefont {Fleming}}, \bibinfo
    {author} {\bibfnamefont {J.}~\bibnamefont {Waszczak}}, \bibinfo {author}
    {\bibfnamefont {R.}~\bibnamefont {Dunn}}, \bibinfo {author} {\bibfnamefont
    {S.}~\bibnamefont {Sunshine}},\ and\ \bibinfo {author} {\bibfnamefont
    {J.~A.}\ \bibnamefont {Ibers}},\ }\href@noop {} {\bibfield  {journal}
    {\bibinfo  {journal} {Journal of the Less Common Metals}\ }\textbf {\bibinfo
    {volume} {116}},\ \bibinfo {pages} {51} (\bibinfo {year} {1986})}\BibitemShut
    {NoStop}%
  \bibitem [{\citenamefont {Lu}\ \emph {et~al.}(2017)\citenamefont {Lu},
    \citenamefont {Kono}, \citenamefont {Larkin}, \citenamefont {Rost},
    \citenamefont {Takayama}, \citenamefont {Boris}, \citenamefont {Keimer},\
    and\ \citenamefont {Takagi}}]{lu2017zero}%
    \BibitemOpen
    \bibfield  {author} {\bibinfo {author} {\bibfnamefont {Y.}~\bibnamefont
    {Lu}}, \bibinfo {author} {\bibfnamefont {H.}~\bibnamefont {Kono}}, \bibinfo
    {author} {\bibfnamefont {T.}~\bibnamefont {Larkin}}, \bibinfo {author}
    {\bibfnamefont {A.}~\bibnamefont {Rost}}, \bibinfo {author} {\bibfnamefont
    {T.}~\bibnamefont {Takayama}}, \bibinfo {author} {\bibfnamefont
    {A.}~\bibnamefont {Boris}}, \bibinfo {author} {\bibfnamefont
    {B.}~\bibnamefont {Keimer}},\ and\ \bibinfo {author} {\bibfnamefont
    {H.}~\bibnamefont {Takagi}},\ }\href@noop {} {\bibfield  {journal} {\bibinfo
    {journal} {Nature communications}\ }\textbf {\bibinfo {volume} {8}},\
    \bibinfo {pages} {14408} (\bibinfo {year} {2017})}\BibitemShut {NoStop}%
  \bibitem [{\citenamefont {Larkin}\ \emph {et~al.}(2017)\citenamefont {Larkin},
    \citenamefont {Yaresko}, \citenamefont {Pr{\"o}pper}, \citenamefont {Kikoin},
    \citenamefont {Lu}, \citenamefont {Takayama}, \citenamefont {Mathis},
    \citenamefont {Rost}, \citenamefont {Takagi}, \citenamefont {Keimer} \emph
    {et~al.}}]{larkin2017giant}%
    \BibitemOpen
    \bibfield  {author} {\bibinfo {author} {\bibfnamefont {T.}~\bibnamefont
    {Larkin}}, \bibinfo {author} {\bibfnamefont {A.}~\bibnamefont {Yaresko}},
    \bibinfo {author} {\bibfnamefont {D.}~\bibnamefont {Pr{\"o}pper}}, \bibinfo
    {author} {\bibfnamefont {K.}~\bibnamefont {Kikoin}}, \bibinfo {author}
    {\bibfnamefont {Y.}~\bibnamefont {Lu}}, \bibinfo {author} {\bibfnamefont
    {T.}~\bibnamefont {Takayama}}, \bibinfo {author} {\bibfnamefont {Y.-L.}\
    \bibnamefont {Mathis}}, \bibinfo {author} {\bibfnamefont {A.}~\bibnamefont
    {Rost}}, \bibinfo {author} {\bibfnamefont {H.}~\bibnamefont {Takagi}},
    \bibinfo {author} {\bibfnamefont {B.}~\bibnamefont {Keimer}}, \emph
    {et~al.},\ }\href@noop {} {\bibfield  {journal} {\bibinfo  {journal}
    {Physical Review B}\ }\textbf {\bibinfo {volume} {95}},\ \bibinfo {pages}
    {195144} (\bibinfo {year} {2017})}\BibitemShut {NoStop}%
  \bibitem [{\citenamefont {Larkin}\ \emph {et~al.}(2018)\citenamefont {Larkin},
    \citenamefont {Dawson}, \citenamefont {H{\"o}ppner}, \citenamefont
    {Takayama}, \citenamefont {Isobe}, \citenamefont {Mathis}, \citenamefont
    {Takagi}, \citenamefont {Keimer},\ and\ \citenamefont
    {Boris}}]{larkin2018infrared}%
    \BibitemOpen
    \bibfield  {author} {\bibinfo {author} {\bibfnamefont {T.}~\bibnamefont
    {Larkin}}, \bibinfo {author} {\bibfnamefont {R.}~\bibnamefont {Dawson}},
    \bibinfo {author} {\bibfnamefont {M.}~\bibnamefont {H{\"o}ppner}}, \bibinfo
    {author} {\bibfnamefont {T.}~\bibnamefont {Takayama}}, \bibinfo {author}
    {\bibfnamefont {M.}~\bibnamefont {Isobe}}, \bibinfo {author} {\bibfnamefont
    {Y.-L.}\ \bibnamefont {Mathis}}, \bibinfo {author} {\bibfnamefont
    {H.}~\bibnamefont {Takagi}}, \bibinfo {author} {\bibfnamefont
    {B.}~\bibnamefont {Keimer}},\ and\ \bibinfo {author} {\bibfnamefont
    {A.}~\bibnamefont {Boris}},\ }\href@noop {} {\bibfield  {journal} {\bibinfo
    {journal} {Physical Review B}\ }\textbf {\bibinfo {volume} {98}},\ \bibinfo
    {pages} {125113} (\bibinfo {year} {2018})}\BibitemShut {NoStop}%
  \bibitem [{\citenamefont {Sunshine}\ and\ \citenamefont
    {Ibers}(1985)}]{sunshine1985structure}%
    \BibitemOpen
    \bibfield  {author} {\bibinfo {author} {\bibfnamefont {S.~A.}\ \bibnamefont
    {Sunshine}}\ and\ \bibinfo {author} {\bibfnamefont {J.~A.}\ \bibnamefont
    {Ibers}},\ }\href@noop {} {\bibfield  {journal} {\bibinfo  {journal}
    {Inorganic Chemistry}\ }\textbf {\bibinfo {volume} {24}},\ \bibinfo {pages}
    {3611} (\bibinfo {year} {1985})}\BibitemShut {NoStop}%
  \bibitem [{\citenamefont {Wakisaka}\ \emph {et~al.}(2012)\citenamefont
    {Wakisaka}, \citenamefont {Sudayama}, \citenamefont {Takubo}, \citenamefont
    {Mizokawa}, \citenamefont {Saini}, \citenamefont {Arita}, \citenamefont
    {Namatame}, \citenamefont {Taniguchi}, \citenamefont {Katayama},
    \citenamefont {Nohara} \emph {et~al.}}]{wakisaka2012photoemission}%
    \BibitemOpen
    \bibfield  {author} {\bibinfo {author} {\bibfnamefont {Y.}~\bibnamefont
    {Wakisaka}}, \bibinfo {author} {\bibfnamefont {T.}~\bibnamefont {Sudayama}},
    \bibinfo {author} {\bibfnamefont {K.}~\bibnamefont {Takubo}}, \bibinfo
    {author} {\bibfnamefont {T.}~\bibnamefont {Mizokawa}}, \bibinfo {author}
    {\bibfnamefont {N.}~\bibnamefont {Saini}}, \bibinfo {author} {\bibfnamefont
    {M.}~\bibnamefont {Arita}}, \bibinfo {author} {\bibfnamefont
    {H.}~\bibnamefont {Namatame}}, \bibinfo {author} {\bibfnamefont
    {M.}~\bibnamefont {Taniguchi}}, \bibinfo {author} {\bibfnamefont
    {N.}~\bibnamefont {Katayama}}, \bibinfo {author} {\bibfnamefont
    {M.}~\bibnamefont {Nohara}}, \emph {et~al.},\ }\href@noop {} {\bibfield
    {journal} {\bibinfo  {journal} {Journal of superconductivity and novel
    magnetism}\ }\textbf {\bibinfo {volume} {25}},\ \bibinfo {pages} {1231}
    (\bibinfo {year} {2012})}\BibitemShut {NoStop}%
  \bibitem [{\citenamefont {Seki}\ \emph {et~al.}(2014)\citenamefont {Seki},
    \citenamefont {Wakisaka}, \citenamefont {Kaneko}, \citenamefont {Toriyama},
    \citenamefont {Konishi}, \citenamefont {Sudayama}, \citenamefont {Saini},
    \citenamefont {Arita}, \citenamefont {Namatame}, \citenamefont {Taniguchi}
    \emph {et~al.}}]{seki2014excitonic}%
    \BibitemOpen
    \bibfield  {author} {\bibinfo {author} {\bibfnamefont {K.}~\bibnamefont
    {Seki}}, \bibinfo {author} {\bibfnamefont {Y.}~\bibnamefont {Wakisaka}},
    \bibinfo {author} {\bibfnamefont {T.}~\bibnamefont {Kaneko}}, \bibinfo
    {author} {\bibfnamefont {T.}~\bibnamefont {Toriyama}}, \bibinfo {author}
    {\bibfnamefont {T.}~\bibnamefont {Konishi}}, \bibinfo {author} {\bibfnamefont
    {T.}~\bibnamefont {Sudayama}}, \bibinfo {author} {\bibfnamefont
    {N.}~\bibnamefont {Saini}}, \bibinfo {author} {\bibfnamefont
    {M.}~\bibnamefont {Arita}}, \bibinfo {author} {\bibfnamefont
    {H.}~\bibnamefont {Namatame}}, \bibinfo {author} {\bibfnamefont
    {M.}~\bibnamefont {Taniguchi}}, \emph {et~al.},\ }\href@noop {} {\bibfield
    {journal} {\bibinfo  {journal} {Physical Review B}\ }\textbf {\bibinfo
    {volume} {90}},\ \bibinfo {pages} {155116} (\bibinfo {year}
    {2014})}\BibitemShut {NoStop}%
  \bibitem [{\citenamefont {Mor}\ \emph {et~al.}(2017)\citenamefont {Mor},
    \citenamefont {Herzog}, \citenamefont {Gole\ifmmode~\check{z}\else
    \v{z}\fi{}}, \citenamefont {Werner}, \citenamefont {Eckstein}, \citenamefont
    {Katayama}, \citenamefont {Nohara}, \citenamefont {Takagi}, \citenamefont
    {Mizokawa}, \citenamefont {Monney},\ and\ \citenamefont
    {St\"ahler}}]{Mor2017}%
    \BibitemOpen
    \bibfield  {author} {\bibinfo {author} {\bibfnamefont {S.}~\bibnamefont
    {Mor}}, \bibinfo {author} {\bibfnamefont {M.}~\bibnamefont {Herzog}},
    \bibinfo {author} {\bibfnamefont {D.}~\bibnamefont
    {Gole\ifmmode~\check{z}\else \v{z}\fi{}}}, \bibinfo {author} {\bibfnamefont
    {P.}~\bibnamefont {Werner}}, \bibinfo {author} {\bibfnamefont
    {M.}~\bibnamefont {Eckstein}}, \bibinfo {author} {\bibfnamefont
    {N.}~\bibnamefont {Katayama}}, \bibinfo {author} {\bibfnamefont
    {M.}~\bibnamefont {Nohara}}, \bibinfo {author} {\bibfnamefont
    {H.}~\bibnamefont {Takagi}}, \bibinfo {author} {\bibfnamefont
    {T.}~\bibnamefont {Mizokawa}}, \bibinfo {author} {\bibfnamefont
    {C.}~\bibnamefont {Monney}},\ and\ \bibinfo {author} {\bibfnamefont
    {J.}~\bibnamefont {St\"ahler}},\ }\href
    {https://doi.org/10.1103/PhysRevLett.119.086401} {\bibfield  {journal}
    {\bibinfo  {journal} {Phys. Rev. Lett.}\ }\textbf {\bibinfo {volume} {119}},\
    \bibinfo {pages} {086401} (\bibinfo {year} {2017})}\BibitemShut {NoStop}%
  \bibitem [{\citenamefont {Mor}\ \emph {et~al.}(2018)\citenamefont {Mor},
    \citenamefont {Herzog}, \citenamefont {Noack}, \citenamefont {Katayama},
    \citenamefont {Nohara}, \citenamefont {Takagi}, \citenamefont {Trunschke},
    \citenamefont {Mizokawa}, \citenamefont {Monney},\ and\ \citenamefont
    {St\"ahler}}]{mor2018}%
    \BibitemOpen
    \bibfield  {author} {\bibinfo {author} {\bibfnamefont {S.}~\bibnamefont
    {Mor}}, \bibinfo {author} {\bibfnamefont {M.}~\bibnamefont {Herzog}},
    \bibinfo {author} {\bibfnamefont {J.}~\bibnamefont {Noack}}, \bibinfo
    {author} {\bibfnamefont {N.}~\bibnamefont {Katayama}}, \bibinfo {author}
    {\bibfnamefont {M.}~\bibnamefont {Nohara}}, \bibinfo {author} {\bibfnamefont
    {H.}~\bibnamefont {Takagi}}, \bibinfo {author} {\bibfnamefont
    {A.}~\bibnamefont {Trunschke}}, \bibinfo {author} {\bibfnamefont
    {T.}~\bibnamefont {Mizokawa}}, \bibinfo {author} {\bibfnamefont
    {C.}~\bibnamefont {Monney}},\ and\ \bibinfo {author} {\bibfnamefont
    {J.}~\bibnamefont {St\"ahler}},\ }\href
    {https://doi.org/10.1103/PhysRevB.97.115154} {\bibfield  {journal} {\bibinfo
    {journal} {Phys. Rev. B}\ }\textbf {\bibinfo {volume} {97}},\ \bibinfo
    {pages} {115154} (\bibinfo {year} {2018})}\BibitemShut {NoStop}%
  \bibitem [{\citenamefont {Okazaki}\ \emph {et~al.}(2018)\citenamefont
    {Okazaki}, \citenamefont {Ogawa}, \citenamefont {Suzuki}, \citenamefont
    {Yamamoto}, \citenamefont {Someya}, \citenamefont {Michimae}, \citenamefont
    {Watanabe}, \citenamefont {Lu}, \citenamefont {Nohara}, \citenamefont
    {Takagi}, \citenamefont {Katayama}, \citenamefont {Sawa}, \citenamefont
    {Fujisawa}, \citenamefont {Kanai}, \citenamefont {Ishii}, \citenamefont
    {Itatani}, \citenamefont {Mizokawa},\ and\ \citenamefont
    {Shin}}]{Okazaki2018}%
    \BibitemOpen
    \bibfield  {author} {\bibinfo {author} {\bibfnamefont {K.}~\bibnamefont
    {Okazaki}}, \bibinfo {author} {\bibfnamefont {Y.}~\bibnamefont {Ogawa}},
    \bibinfo {author} {\bibfnamefont {T.}~\bibnamefont {Suzuki}}, \bibinfo
    {author} {\bibfnamefont {T.}~\bibnamefont {Yamamoto}}, \bibinfo {author}
    {\bibfnamefont {T.}~\bibnamefont {Someya}}, \bibinfo {author} {\bibfnamefont
    {S.}~\bibnamefont {Michimae}}, \bibinfo {author} {\bibfnamefont
    {M.}~\bibnamefont {Watanabe}}, \bibinfo {author} {\bibfnamefont
    {Y.}~\bibnamefont {Lu}}, \bibinfo {author} {\bibfnamefont {M.}~\bibnamefont
    {Nohara}}, \bibinfo {author} {\bibfnamefont {H.}~\bibnamefont {Takagi}},
    \bibinfo {author} {\bibfnamefont {N.}~\bibnamefont {Katayama}}, \bibinfo
    {author} {\bibfnamefont {H.}~\bibnamefont {Sawa}}, \bibinfo {author}
    {\bibfnamefont {M.}~\bibnamefont {Fujisawa}}, \bibinfo {author}
    {\bibfnamefont {T.}~\bibnamefont {Kanai}}, \bibinfo {author} {\bibfnamefont
    {N.}~\bibnamefont {Ishii}}, \bibinfo {author} {\bibfnamefont
    {J.}~\bibnamefont {Itatani}}, \bibinfo {author} {\bibfnamefont
    {T.}~\bibnamefont {Mizokawa}},\ and\ \bibinfo {author} {\bibfnamefont
    {S.}~\bibnamefont {Shin}},\ }\bibfield  {journal} {\bibinfo  {journal}
    {Nature Communications}\ }\textbf {\bibinfo {volume} {9}},\ \href
    {https://doi.org/10.1038/s41467-018-06801-1} {10.1038/s41467-018-06801-1}
    (\bibinfo {year} {2018})\BibitemShut {NoStop}%
  \bibitem [{\citenamefont {Bretscher}\ \emph
    {et~al.}(2021{\natexlab{a}})\citenamefont {Bretscher}, \citenamefont
    {Andrich}, \citenamefont {Telang}, \citenamefont {Singh}, \citenamefont
    {Harnagea}, \citenamefont {Sood},\ and\ \citenamefont {Rao}}]{Bretscher2021}%
    \BibitemOpen
    \bibfield  {author} {\bibinfo {author} {\bibfnamefont {H.~M.}\ \bibnamefont
    {Bretscher}}, \bibinfo {author} {\bibfnamefont {P.}~\bibnamefont {Andrich}},
    \bibinfo {author} {\bibfnamefont {P.}~\bibnamefont {Telang}}, \bibinfo
    {author} {\bibfnamefont {A.}~\bibnamefont {Singh}}, \bibinfo {author}
    {\bibfnamefont {L.}~\bibnamefont {Harnagea}}, \bibinfo {author}
    {\bibfnamefont {A.~K.}\ \bibnamefont {Sood}},\ and\ \bibinfo {author}
    {\bibfnamefont {A.}~\bibnamefont {Rao}},\ }\bibfield  {journal} {\bibinfo
    {journal} {Nature Communications}\ }\textbf {\bibinfo {volume} {12}},\ \href
    {https://doi.org/10.1038/s41467-021-21929-3} {10.1038/s41467-021-21929-3}
    (\bibinfo {year} {2021}{\natexlab{a}})\BibitemShut {NoStop}%
  \bibitem [{\citenamefont {Bretscher}\ \emph
    {et~al.}(2021{\natexlab{b}})\citenamefont {Bretscher}, \citenamefont
    {Andrich}, \citenamefont {Murakami}, \citenamefont {Golež}, \citenamefont
    {Remez}, \citenamefont {Telang}, \citenamefont {Singh}, \citenamefont
    {Harnagea}, \citenamefont {Cooper}, \citenamefont {Millis}, \citenamefont
    {Werner}, \citenamefont {Sood},\ and\ \citenamefont {Rao}}]{Bretscher2021b}%
    \BibitemOpen
    \bibfield  {author} {\bibinfo {author} {\bibfnamefont {H.~M.}\ \bibnamefont
    {Bretscher}}, \bibinfo {author} {\bibfnamefont {P.}~\bibnamefont {Andrich}},
    \bibinfo {author} {\bibfnamefont {Y.}~\bibnamefont {Murakami}}, \bibinfo
    {author} {\bibfnamefont {D.}~\bibnamefont {Golež}}, \bibinfo {author}
    {\bibfnamefont {B.}~\bibnamefont {Remez}}, \bibinfo {author} {\bibfnamefont
    {P.}~\bibnamefont {Telang}}, \bibinfo {author} {\bibfnamefont
    {A.}~\bibnamefont {Singh}}, \bibinfo {author} {\bibfnamefont
    {L.}~\bibnamefont {Harnagea}}, \bibinfo {author} {\bibfnamefont {N.~R.}\
    \bibnamefont {Cooper}}, \bibinfo {author} {\bibfnamefont {A.~J.}\
    \bibnamefont {Millis}}, \bibinfo {author} {\bibfnamefont {P.}~\bibnamefont
    {Werner}}, \bibinfo {author} {\bibfnamefont {A.~K.}\ \bibnamefont {Sood}},\
    and\ \bibinfo {author} {\bibfnamefont {A.}~\bibnamefont {Rao}},\ }\bibfield
    {journal} {\bibinfo  {journal} {Science Advances}\ }\textbf {\bibinfo
    {volume} {7}},\ \href {https://doi.org/10.1126/sciadv.abd6147}
    {10.1126/sciadv.abd6147} (\bibinfo {year} {2021}{\natexlab{b}})\BibitemShut
    {NoStop}%
  \bibitem [{\citenamefont {Saha}\ \emph {et~al.}(2021)\citenamefont {Saha},
    \citenamefont {Gole\ifmmode~\check{z}\else \v{z}\fi{}}, \citenamefont
    {De~Ninno}, \citenamefont {Mravlje}, \citenamefont {Murakami}, \citenamefont
    {Ressel}, \citenamefont {Stupar},\ and\ \citenamefont
    {Ribi\ifmmode~\check{c}\else \v{c}\fi{}}}]{Saha2021}%
    \BibitemOpen
    \bibfield  {author} {\bibinfo {author} {\bibfnamefont {T.}~\bibnamefont
    {Saha}}, \bibinfo {author} {\bibfnamefont {D.}~\bibnamefont
    {Gole\ifmmode~\check{z}\else \v{z}\fi{}}}, \bibinfo {author} {\bibfnamefont
    {G.}~\bibnamefont {De~Ninno}}, \bibinfo {author} {\bibfnamefont
    {J.}~\bibnamefont {Mravlje}}, \bibinfo {author} {\bibfnamefont
    {Y.}~\bibnamefont {Murakami}}, \bibinfo {author} {\bibfnamefont
    {B.}~\bibnamefont {Ressel}}, \bibinfo {author} {\bibfnamefont
    {M.}~\bibnamefont {Stupar}},\ and\ \bibinfo {author} {\bibfnamefont {P.~c.
    v.~R.}\ \bibnamefont {Ribi\ifmmode~\check{c}\else \v{c}\fi{}}},\ }\href
    {https://doi.org/10.1103/PhysRevB.103.144304} {\bibfield  {journal} {\bibinfo
     {journal} {Phys. Rev. B}\ }\textbf {\bibinfo {volume} {103}},\ \bibinfo
    {pages} {144304} (\bibinfo {year} {2021})}\BibitemShut {NoStop}%
  \bibitem [{\citenamefont {Geng}\ \emph {et~al.}(2024)\citenamefont {Geng},
    \citenamefont {Liu}, \citenamefont {Zhang}, \citenamefont {Golež},\ and\
    \citenamefont {Peng}}]{geng2024}%
    \BibitemOpen
    \bibfield  {author} {\bibinfo {author} {\bibfnamefont {L.}~\bibnamefont
    {Geng}}, \bibinfo {author} {\bibfnamefont {X.}~\bibnamefont {Liu}}, \bibinfo
    {author} {\bibfnamefont {J.}~\bibnamefont {Zhang}}, \bibinfo {author}
    {\bibfnamefont {D.}~\bibnamefont {Golež}},\ and\ \bibinfo {author}
    {\bibfnamefont {L.-Y.}\ \bibnamefont {Peng}},\ }\href@noop {} {\bibinfo
    {title} {Anomalous photo-induced band renormalization in correlated
    materials: Case study of ta$_2$nise$_5$}} (\bibinfo {year} {2024}),\ \Eprint
    {https://arxiv.org/abs/2401.16988} {arXiv:2401.16988 [cond-mat.str-el]}
    \BibitemShut {NoStop}%
  \bibitem [{\citenamefont {Nakano}\ \emph {et~al.}(2019)\citenamefont {Nakano},
    \citenamefont {Nagai}, \citenamefont {Katayama}, \citenamefont {Sawa},
    \citenamefont {Taniguchi},\ and\ \citenamefont
    {Terasaki}}]{nakano2019exciton}%
    \BibitemOpen
    \bibfield  {author} {\bibinfo {author} {\bibfnamefont {A.}~\bibnamefont
    {Nakano}}, \bibinfo {author} {\bibfnamefont {T.}~\bibnamefont {Nagai}},
    \bibinfo {author} {\bibfnamefont {N.}~\bibnamefont {Katayama}}, \bibinfo
    {author} {\bibfnamefont {H.}~\bibnamefont {Sawa}}, \bibinfo {author}
    {\bibfnamefont {H.}~\bibnamefont {Taniguchi}},\ and\ \bibinfo {author}
    {\bibfnamefont {I.}~\bibnamefont {Terasaki}},\ }\href@noop {} {\bibfield
    {journal} {\bibinfo  {journal} {Journal of the Physical Society of Japan}\
    }\textbf {\bibinfo {volume} {88}},\ \bibinfo {pages} {113706} (\bibinfo
    {year} {2019})}\BibitemShut {NoStop}%
  \bibitem [{\citenamefont {Kaneko}\ \emph {et~al.}(2012)\citenamefont {Kaneko},
    \citenamefont {Toriyama}, \citenamefont {Konishi},\ and\ \citenamefont
    {Ohta}}]{kaneko2012electronic}%
    \BibitemOpen
    \bibfield  {author} {\bibinfo {author} {\bibfnamefont {T.}~\bibnamefont
    {Kaneko}}, \bibinfo {author} {\bibfnamefont {T.}~\bibnamefont {Toriyama}},
    \bibinfo {author} {\bibfnamefont {T.}~\bibnamefont {Konishi}},\ and\ \bibinfo
    {author} {\bibfnamefont {Y.}~\bibnamefont {Ohta}},\ }in\ \href@noop {} {\emph
    {\bibinfo {booktitle} {Journal of Physics: Conference Series}}},\ Vol.\
    \bibinfo {volume} {400}\ (\bibinfo {organization} {IOP Publishing},\ \bibinfo
    {year} {2012})\ p.\ \bibinfo {pages} {032035}\BibitemShut {NoStop}%
  \bibitem [{\citenamefont {Kaneko}\ \emph {et~al.}(2013)\citenamefont {Kaneko},
    \citenamefont {Toriyama}, \citenamefont {Konishi},\ and\ \citenamefont
    {Ohta}}]{kaneko2013orthorhombic}%
    \BibitemOpen
    \bibfield  {author} {\bibinfo {author} {\bibfnamefont {T.}~\bibnamefont
    {Kaneko}}, \bibinfo {author} {\bibfnamefont {T.}~\bibnamefont {Toriyama}},
    \bibinfo {author} {\bibfnamefont {T.}~\bibnamefont {Konishi}},\ and\ \bibinfo
    {author} {\bibfnamefont {Y.}~\bibnamefont {Ohta}},\ }\href@noop {} {\bibfield
     {journal} {\bibinfo  {journal} {Physical Review B}\ }\textbf {\bibinfo
    {volume} {87}},\ \bibinfo {pages} {035121} (\bibinfo {year}
    {2013})}\BibitemShut {NoStop}%
  \bibitem [{\citenamefont {Subedi}(2020)}]{subedi2020orthorhombic}%
    \BibitemOpen
    \bibfield  {author} {\bibinfo {author} {\bibfnamefont {A.}~\bibnamefont
    {Subedi}},\ }\href@noop {} {\bibfield  {journal} {\bibinfo  {journal}
    {Physical Review Materials}\ }\textbf {\bibinfo {volume} {4}},\ \bibinfo
    {pages} {083601} (\bibinfo {year} {2020})}\BibitemShut {NoStop}%
  \bibitem [{\citenamefont {Watson}\ \emph {et~al.}(2020)\citenamefont {Watson},
    \citenamefont {Markovi{\'c}}, \citenamefont {Morales}, \citenamefont
    {Le~F{\`e}vre}, \citenamefont {Merz}, \citenamefont {Haghighirad},\ and\
    \citenamefont {King}}]{watson2020band}%
    \BibitemOpen
    \bibfield  {author} {\bibinfo {author} {\bibfnamefont {M.~D.}\ \bibnamefont
    {Watson}}, \bibinfo {author} {\bibfnamefont {I.}~\bibnamefont
    {Markovi{\'c}}}, \bibinfo {author} {\bibfnamefont {E.~A.}\ \bibnamefont
    {Morales}}, \bibinfo {author} {\bibfnamefont {P.}~\bibnamefont
    {Le~F{\`e}vre}}, \bibinfo {author} {\bibfnamefont {M.}~\bibnamefont {Merz}},
    \bibinfo {author} {\bibfnamefont {A.~A.}\ \bibnamefont {Haghighirad}},\ and\
    \bibinfo {author} {\bibfnamefont {P.~D.}\ \bibnamefont {King}},\ }\href@noop
    {} {\bibfield  {journal} {\bibinfo  {journal} {Physical Review Research}\
    }\textbf {\bibinfo {volume} {2}},\ \bibinfo {pages} {013236} (\bibinfo {year}
    {2020})}\BibitemShut {NoStop}%
  \bibitem [{\citenamefont {Windg{\"a}tter}\ \emph {et~al.}(2021)\citenamefont
    {Windg{\"a}tter}, \citenamefont {R{\"o}sner}, \citenamefont {Mazza},
    \citenamefont {H{\"u}bener}, \citenamefont {Georges}, \citenamefont {Millis},
    \citenamefont {Latini},\ and\ \citenamefont {Rubio}}]{windgatter2021common}%
    \BibitemOpen
    \bibfield  {author} {\bibinfo {author} {\bibfnamefont {L.}~\bibnamefont
    {Windg{\"a}tter}}, \bibinfo {author} {\bibfnamefont {M.}~\bibnamefont
    {R{\"o}sner}}, \bibinfo {author} {\bibfnamefont {G.}~\bibnamefont {Mazza}},
    \bibinfo {author} {\bibfnamefont {H.}~\bibnamefont {H{\"u}bener}}, \bibinfo
    {author} {\bibfnamefont {A.}~\bibnamefont {Georges}}, \bibinfo {author}
    {\bibfnamefont {A.~J.}\ \bibnamefont {Millis}}, \bibinfo {author}
    {\bibfnamefont {S.}~\bibnamefont {Latini}},\ and\ \bibinfo {author}
    {\bibfnamefont {A.}~\bibnamefont {Rubio}},\ }\href@noop {} {\bibfield
    {journal} {\bibinfo  {journal} {npj Computational Materials}\ }\textbf
    {\bibinfo {volume} {7}},\ \bibinfo {pages} {210} (\bibinfo {year}
    {2021})}\BibitemShut {NoStop}%
  \bibitem [{\citenamefont {Baldini}\ \emph {et~al.}(2023)\citenamefont
    {Baldini}, \citenamefont {Zong}, \citenamefont {Choi}, \citenamefont {Lee},
    \citenamefont {Michael}, \citenamefont {Windgaetter}, \citenamefont {Mazin},
    \citenamefont {Latini}, \citenamefont {Azoury}, \citenamefont {Lv} \emph
    {et~al.}}]{baldini2023spontaneous}%
    \BibitemOpen
    \bibfield  {author} {\bibinfo {author} {\bibfnamefont {E.}~\bibnamefont
    {Baldini}}, \bibinfo {author} {\bibfnamefont {A.}~\bibnamefont {Zong}},
    \bibinfo {author} {\bibfnamefont {D.}~\bibnamefont {Choi}}, \bibinfo {author}
    {\bibfnamefont {C.}~\bibnamefont {Lee}}, \bibinfo {author} {\bibfnamefont
    {M.~H.}\ \bibnamefont {Michael}}, \bibinfo {author} {\bibfnamefont
    {L.}~\bibnamefont {Windgaetter}}, \bibinfo {author} {\bibfnamefont {I.~I.}\
    \bibnamefont {Mazin}}, \bibinfo {author} {\bibfnamefont {S.}~\bibnamefont
    {Latini}}, \bibinfo {author} {\bibfnamefont {D.}~\bibnamefont {Azoury}},
    \bibinfo {author} {\bibfnamefont {B.}~\bibnamefont {Lv}}, \emph {et~al.},\
    }\href@noop {} {\bibfield  {journal} {\bibinfo  {journal} {Proceedings of the
    National Academy of Sciences}\ }\textbf {\bibinfo {volume} {120}},\ \bibinfo
    {pages} {e2221688120} (\bibinfo {year} {2023})}\BibitemShut {NoStop}%
  \bibitem [{\citenamefont {Ye}\ \emph {et~al.}(2021)\citenamefont {Ye},
    \citenamefont {Volkov}, \citenamefont {Lohani}, \citenamefont {Feldman},
    \citenamefont {Kim}, \citenamefont {Kanigel},\ and\ \citenamefont
    {Blumberg}}]{ye2021lattice}%
    \BibitemOpen
    \bibfield  {author} {\bibinfo {author} {\bibfnamefont {M.}~\bibnamefont
    {Ye}}, \bibinfo {author} {\bibfnamefont {P.~A.}\ \bibnamefont {Volkov}},
    \bibinfo {author} {\bibfnamefont {H.}~\bibnamefont {Lohani}}, \bibinfo
    {author} {\bibfnamefont {I.}~\bibnamefont {Feldman}}, \bibinfo {author}
    {\bibfnamefont {M.}~\bibnamefont {Kim}}, \bibinfo {author} {\bibfnamefont
    {A.}~\bibnamefont {Kanigel}},\ and\ \bibinfo {author} {\bibfnamefont
    {G.}~\bibnamefont {Blumberg}},\ }\href
    {https://doi.org/10.1103/PhysRevB.104.045102} {\bibfield  {journal} {\bibinfo
     {journal} {Phys. Rev. B}\ }\textbf {\bibinfo {volume} {104}},\ \bibinfo
    {pages} {045102} (\bibinfo {year} {2021})}\BibitemShut {NoStop}%
  \bibitem [{\citenamefont {Gole\ifmmode~\check{z}\else \v{z}\fi{}}\ \emph
    {et~al.}(2022)\citenamefont {Gole\ifmmode~\check{z}\else \v{z}\fi{}},
    \citenamefont {Dufresne}, \citenamefont {Kim}, \citenamefont {Boschini},
    \citenamefont {Chu}, \citenamefont {Murakami}, \citenamefont {Levy},
    \citenamefont {Mills}, \citenamefont {Zhdanovich}, \citenamefont {Isobe},
    \citenamefont {Takagi}, \citenamefont {Kaiser}, \citenamefont {Werner},
    \citenamefont {Jones}, \citenamefont {Georges}, \citenamefont {Damascelli},\
    and\ \citenamefont {Millis}}]{golez2022}%
    \BibitemOpen
    \bibfield  {author} {\bibinfo {author} {\bibfnamefont {D.}~\bibnamefont
    {Gole\ifmmode~\check{z}\else \v{z}\fi{}}}, \bibinfo {author} {\bibfnamefont
    {S.~K.~Y.}\ \bibnamefont {Dufresne}}, \bibinfo {author} {\bibfnamefont
    {M.-J.}\ \bibnamefont {Kim}}, \bibinfo {author} {\bibfnamefont
    {F.}~\bibnamefont {Boschini}}, \bibinfo {author} {\bibfnamefont
    {H.}~\bibnamefont {Chu}}, \bibinfo {author} {\bibfnamefont {Y.}~\bibnamefont
    {Murakami}}, \bibinfo {author} {\bibfnamefont {G.}~\bibnamefont {Levy}},
    \bibinfo {author} {\bibfnamefont {A.~K.}\ \bibnamefont {Mills}}, \bibinfo
    {author} {\bibfnamefont {S.}~\bibnamefont {Zhdanovich}}, \bibinfo {author}
    {\bibfnamefont {M.}~\bibnamefont {Isobe}}, \bibinfo {author} {\bibfnamefont
    {H.}~\bibnamefont {Takagi}}, \bibinfo {author} {\bibfnamefont
    {S.}~\bibnamefont {Kaiser}}, \bibinfo {author} {\bibfnamefont
    {P.}~\bibnamefont {Werner}}, \bibinfo {author} {\bibfnamefont {D.~J.}\
    \bibnamefont {Jones}}, \bibinfo {author} {\bibfnamefont {A.}~\bibnamefont
    {Georges}}, \bibinfo {author} {\bibfnamefont {A.}~\bibnamefont
    {Damascelli}},\ and\ \bibinfo {author} {\bibfnamefont {A.~J.}\ \bibnamefont
    {Millis}},\ }\href {https://doi.org/10.1103/PhysRevB.106.L121106} {\bibfield
    {journal} {\bibinfo  {journal} {Phys. Rev. B}\ }\textbf {\bibinfo {volume}
    {106}},\ \bibinfo {pages} {L121106} (\bibinfo {year} {2022})}\BibitemShut
    {NoStop}%
  \bibitem [{\citenamefont {Kim}\ \emph {et~al.}(2020)\citenamefont {Kim},
    \citenamefont {Schulz}, \citenamefont {Takayama}, \citenamefont {Isobe},
    \citenamefont {Takagi},\ and\ \citenamefont {Kaiser}}]{Kim2020}%
    \BibitemOpen
    \bibfield  {author} {\bibinfo {author} {\bibfnamefont {M.-J.}\ \bibnamefont
    {Kim}}, \bibinfo {author} {\bibfnamefont {A.}~\bibnamefont {Schulz}},
    \bibinfo {author} {\bibfnamefont {T.}~\bibnamefont {Takayama}}, \bibinfo
    {author} {\bibfnamefont {M.}~\bibnamefont {Isobe}}, \bibinfo {author}
    {\bibfnamefont {H.}~\bibnamefont {Takagi}},\ and\ \bibinfo {author}
    {\bibfnamefont {S.}~\bibnamefont {Kaiser}},\ }\href
    {https://doi.org/10.1103/PhysRevResearch.2.042039} {\bibfield  {journal}
    {\bibinfo  {journal} {Phys. Rev. Res.}\ }\textbf {\bibinfo {volume} {2}},\
    \bibinfo {pages} {042039} (\bibinfo {year} {2020})}\BibitemShut {NoStop}%
  \bibitem [{\citenamefont {Kim}\ \emph {et~al.}(2021)\citenamefont {Kim},
    \citenamefont {Kim}, \citenamefont {Kim}, \citenamefont {Kwon}, \citenamefont
    {Kim},\ and\ \citenamefont {Kim}}]{kim2021direct}%
    \BibitemOpen
    \bibfield  {author} {\bibinfo {author} {\bibfnamefont {K.}~\bibnamefont
    {Kim}}, \bibinfo {author} {\bibfnamefont {H.}~\bibnamefont {Kim}}, \bibinfo
    {author} {\bibfnamefont {J.}~\bibnamefont {Kim}}, \bibinfo {author}
    {\bibfnamefont {C.}~\bibnamefont {Kwon}}, \bibinfo {author} {\bibfnamefont
    {J.~S.}\ \bibnamefont {Kim}},\ and\ \bibinfo {author} {\bibfnamefont
    {B.}~\bibnamefont {Kim}},\ }\href@noop {} {\bibfield  {journal} {\bibinfo
    {journal} {Nature communications}\ }\textbf {\bibinfo {volume} {12}},\
    \bibinfo {pages} {1969} (\bibinfo {year} {2021})}\BibitemShut {NoStop}%
  \bibitem [{\citenamefont {Volkov}\ \emph
    {et~al.}(2021{\natexlab{a}})\citenamefont {Volkov}, \citenamefont {Ye},
    \citenamefont {Lohani}, \citenamefont {Feldman}, \citenamefont {Kanigel},\
    and\ \citenamefont {Blumberg}}]{Volkov2021}%
    \BibitemOpen
    \bibfield  {author} {\bibinfo {author} {\bibfnamefont {P.~A.}\ \bibnamefont
    {Volkov}}, \bibinfo {author} {\bibfnamefont {M.}~\bibnamefont {Ye}}, \bibinfo
    {author} {\bibfnamefont {H.}~\bibnamefont {Lohani}}, \bibinfo {author}
    {\bibfnamefont {I.}~\bibnamefont {Feldman}}, \bibinfo {author} {\bibfnamefont
    {A.}~\bibnamefont {Kanigel}},\ and\ \bibinfo {author} {\bibfnamefont
    {G.}~\bibnamefont {Blumberg}},\ }\bibfield  {journal} {\bibinfo  {journal}
    {npj Quantum Materials}\ }\textbf {\bibinfo {volume} {6}},\ \href
    {https://doi.org/10.1038/s41535-021-00351-4} {10.1038/s41535-021-00351-4}
    (\bibinfo {year} {2021}{\natexlab{a}})\BibitemShut {NoStop}%
  \bibitem [{\citenamefont {Volkov}\ \emph
    {et~al.}(2021{\natexlab{b}})\citenamefont {Volkov}, \citenamefont {Ye},
    \citenamefont {Lohani}, \citenamefont {Feldman}, \citenamefont {Kanigel},\
    and\ \citenamefont {Blumberg}}]{volkov2021failed}%
    \BibitemOpen
    \bibfield  {author} {\bibinfo {author} {\bibfnamefont {P.~A.}\ \bibnamefont
    {Volkov}}, \bibinfo {author} {\bibfnamefont {M.}~\bibnamefont {Ye}}, \bibinfo
    {author} {\bibfnamefont {H.}~\bibnamefont {Lohani}}, \bibinfo {author}
    {\bibfnamefont {I.}~\bibnamefont {Feldman}}, \bibinfo {author} {\bibfnamefont
    {A.}~\bibnamefont {Kanigel}},\ and\ \bibinfo {author} {\bibfnamefont
    {G.}~\bibnamefont {Blumberg}},\ }\href@noop {} {\bibfield  {journal}
    {\bibinfo  {journal} {Physical Review B}\ }\textbf {\bibinfo {volume}
    {104}},\ \bibinfo {pages} {L241103} (\bibinfo {year}
    {2021}{\natexlab{b}})}\BibitemShut {NoStop}%
  \bibitem [{\citenamefont {Yan}\ \emph {et~al.}(2019)\citenamefont {Yan},
    \citenamefont {Xiao}, \citenamefont {Luo}, \citenamefont {Lv}, \citenamefont
    {Zhang}, \citenamefont {Sun}, \citenamefont {Tong}, \citenamefont {Lu},
    \citenamefont {Song}, \citenamefont {Zhu} \emph {et~al.}}]{yan2019strong}%
    \BibitemOpen
    \bibfield  {author} {\bibinfo {author} {\bibfnamefont {J.}~\bibnamefont
    {Yan}}, \bibinfo {author} {\bibfnamefont {R.}~\bibnamefont {Xiao}}, \bibinfo
    {author} {\bibfnamefont {X.}~\bibnamefont {Luo}}, \bibinfo {author}
    {\bibfnamefont {H.}~\bibnamefont {Lv}}, \bibinfo {author} {\bibfnamefont
    {R.}~\bibnamefont {Zhang}}, \bibinfo {author} {\bibfnamefont
    {Y.}~\bibnamefont {Sun}}, \bibinfo {author} {\bibfnamefont {P.}~\bibnamefont
    {Tong}}, \bibinfo {author} {\bibfnamefont {W.}~\bibnamefont {Lu}}, \bibinfo
    {author} {\bibfnamefont {W.}~\bibnamefont {Song}}, \bibinfo {author}
    {\bibfnamefont {X.}~\bibnamefont {Zhu}}, \emph {et~al.},\ }\href@noop {}
    {\bibfield  {journal} {\bibinfo  {journal} {Inorganic Chemistry}\ }\textbf
    {\bibinfo {volume} {58}},\ \bibinfo {pages} {9036} (\bibinfo {year}
    {2019})}\BibitemShut {NoStop}%
  \bibitem [{\citenamefont {Katsumi}\ \emph {et~al.}(2023)\citenamefont
    {Katsumi}, \citenamefont {Alekhin}, \citenamefont {Souliou}, \citenamefont
    {Merz}, \citenamefont {Haghighirad}, \citenamefont {Le~Tacon}, \citenamefont
    {Houver}, \citenamefont {Cazayous}, \citenamefont {Sacuto},\ and\
    \citenamefont {Gallais}}]{katsumi2023disentangling}%
    \BibitemOpen
    \bibfield  {author} {\bibinfo {author} {\bibfnamefont {K.}~\bibnamefont
    {Katsumi}}, \bibinfo {author} {\bibfnamefont {A.}~\bibnamefont {Alekhin}},
    \bibinfo {author} {\bibfnamefont {S.-M.}\ \bibnamefont {Souliou}}, \bibinfo
    {author} {\bibfnamefont {M.}~\bibnamefont {Merz}}, \bibinfo {author}
    {\bibfnamefont {A.-A.}\ \bibnamefont {Haghighirad}}, \bibinfo {author}
    {\bibfnamefont {M.}~\bibnamefont {Le~Tacon}}, \bibinfo {author}
    {\bibfnamefont {S.}~\bibnamefont {Houver}}, \bibinfo {author} {\bibfnamefont
    {M.}~\bibnamefont {Cazayous}}, \bibinfo {author} {\bibfnamefont
    {A.}~\bibnamefont {Sacuto}},\ and\ \bibinfo {author} {\bibfnamefont
    {Y.}~\bibnamefont {Gallais}},\ }\href@noop {} {\bibfield  {journal} {\bibinfo
     {journal} {Physical Review Letters}\ }\textbf {\bibinfo {volume} {130}},\
    \bibinfo {pages} {106904} (\bibinfo {year} {2023})}\BibitemShut {NoStop}%
  \bibitem [{\citenamefont {Tang}\ \emph {et~al.}(2020)\citenamefont {Tang},
    \citenamefont {Wang}, \citenamefont {Duan}, \citenamefont {Yang},
    \citenamefont {Huang}, \citenamefont {Guo}, \citenamefont {Qian},\ and\
    \citenamefont {Zhang}}]{tang2020non}%
    \BibitemOpen
    \bibfield  {author} {\bibinfo {author} {\bibfnamefont {T.}~\bibnamefont
    {Tang}}, \bibinfo {author} {\bibfnamefont {H.}~\bibnamefont {Wang}}, \bibinfo
    {author} {\bibfnamefont {S.}~\bibnamefont {Duan}}, \bibinfo {author}
    {\bibfnamefont {Y.}~\bibnamefont {Yang}}, \bibinfo {author} {\bibfnamefont
    {C.}~\bibnamefont {Huang}}, \bibinfo {author} {\bibfnamefont
    {Y.}~\bibnamefont {Guo}}, \bibinfo {author} {\bibfnamefont {D.}~\bibnamefont
    {Qian}},\ and\ \bibinfo {author} {\bibfnamefont {W.}~\bibnamefont {Zhang}},\
    }\href@noop {} {\bibfield  {journal} {\bibinfo  {journal} {Physical Review
    B}\ }\textbf {\bibinfo {volume} {101}},\ \bibinfo {pages} {235148} (\bibinfo
    {year} {2020})}\BibitemShut {NoStop}%
  \bibitem [{\citenamefont {Werdehausen}\ \emph {et~al.}(2018)\citenamefont
    {Werdehausen}, \citenamefont {Takayama}, \citenamefont {H\"{o}ppner},
    \citenamefont {Albrecht}, \citenamefont {Rost}, \citenamefont {Lu},
    \citenamefont {Manske}, \citenamefont {Takagi},\ and\ \citenamefont
    {Kaiser}}]{werdehausen2018coherent}%
    \BibitemOpen
    \bibfield  {author} {\bibinfo {author} {\bibfnamefont {D.}~\bibnamefont
    {Werdehausen}}, \bibinfo {author} {\bibfnamefont {T.}~\bibnamefont
    {Takayama}}, \bibinfo {author} {\bibfnamefont {M.}~\bibnamefont
    {H\"{o}ppner}}, \bibinfo {author} {\bibfnamefont {G.}~\bibnamefont
    {Albrecht}}, \bibinfo {author} {\bibfnamefont {A.~W.}\ \bibnamefont {Rost}},
    \bibinfo {author} {\bibfnamefont {Y.}~\bibnamefont {Lu}}, \bibinfo {author}
    {\bibfnamefont {D.}~\bibnamefont {Manske}}, \bibinfo {author} {\bibfnamefont
    {H.}~\bibnamefont {Takagi}},\ and\ \bibinfo {author} {\bibfnamefont
    {S.}~\bibnamefont {Kaiser}},\ }\bibfield  {journal} {\bibinfo  {journal}
    {Science Advances}\ }\textbf {\bibinfo {volume} {4}},\ \href
    {https://doi.org/10.1126/sciadv.aap8652} {10.1126/sciadv.aap8652} (\bibinfo
    {year} {2018})\BibitemShut {NoStop}%
  \bibitem [{\citenamefont {Haque}\ \emph {et~al.}(2024)\citenamefont {Haque},
    \citenamefont {Michael}, \citenamefont {Zhu}, \citenamefont {Zhang},
    \citenamefont {Windg\"{a}tter}, \citenamefont {Latini}, \citenamefont
    {Wakefield}, \citenamefont {Zhang}, \citenamefont {Zhang}, \citenamefont
    {Rubio}, \citenamefont {Checkelsky}, \citenamefont {Demler},\ and\
    \citenamefont {Averitt}}]{Haque2024}%
    \BibitemOpen
    \bibfield  {author} {\bibinfo {author} {\bibfnamefont {S.~R.~U.}\
    \bibnamefont {Haque}}, \bibinfo {author} {\bibfnamefont {M.~H.}\ \bibnamefont
    {Michael}}, \bibinfo {author} {\bibfnamefont {J.}~\bibnamefont {Zhu}},
    \bibinfo {author} {\bibfnamefont {Y.}~\bibnamefont {Zhang}}, \bibinfo
    {author} {\bibfnamefont {L.}~\bibnamefont {Windg\"{a}tter}}, \bibinfo
    {author} {\bibfnamefont {S.}~\bibnamefont {Latini}}, \bibinfo {author}
    {\bibfnamefont {J.~P.}\ \bibnamefont {Wakefield}}, \bibinfo {author}
    {\bibfnamefont {G.-F.}\ \bibnamefont {Zhang}}, \bibinfo {author}
    {\bibfnamefont {J.}~\bibnamefont {Zhang}}, \bibinfo {author} {\bibfnamefont
    {A.}~\bibnamefont {Rubio}}, \bibinfo {author} {\bibfnamefont {J.~G.}\
    \bibnamefont {Checkelsky}}, \bibinfo {author} {\bibfnamefont
    {E.}~\bibnamefont {Demler}},\ and\ \bibinfo {author} {\bibfnamefont {R.~D.}\
    \bibnamefont {Averitt}},\ }\bibfield  {journal} {\bibinfo  {journal} {Nature
    Materials}\ }\href {https://doi.org/10.1038/s41563-023-01755-2}
    {10.1038/s41563-023-01755-2} (\bibinfo {year} {2024})\BibitemShut {NoStop}%
  \bibitem [{\citenamefont {Sugimoto}\ \emph {et~al.}(2016)\citenamefont
    {Sugimoto}, \citenamefont {Kaneko},\ and\ \citenamefont
    {Ohta}}]{sugimoto2016}%
    \BibitemOpen
    \bibfield  {author} {\bibinfo {author} {\bibfnamefont {K.}~\bibnamefont
    {Sugimoto}}, \bibinfo {author} {\bibfnamefont {T.}~\bibnamefont {Kaneko}},\
    and\ \bibinfo {author} {\bibfnamefont {Y.}~\bibnamefont {Ohta}},\ }\href@noop
    {} {\bibfield  {journal} {\bibinfo  {journal} {Physical Review B}\ }\textbf
    {\bibinfo {volume} {93}},\ \bibinfo {pages} {041105} (\bibinfo {year}
    {2016})}\BibitemShut {NoStop}%
  \bibitem [{\citenamefont {Murakami}\ \emph {et~al.}(2020)\citenamefont
    {Murakami}, \citenamefont {Gole{\v{z}}}, \citenamefont {Kaneko},
    \citenamefont {Koga}, \citenamefont {Millis},\ and\ \citenamefont
    {Werner}}]{murakami2020collective}%
    \BibitemOpen
    \bibfield  {author} {\bibinfo {author} {\bibfnamefont {Y.}~\bibnamefont
    {Murakami}}, \bibinfo {author} {\bibfnamefont {D.}~\bibnamefont
    {Gole{\v{z}}}}, \bibinfo {author} {\bibfnamefont {T.}~\bibnamefont {Kaneko}},
    \bibinfo {author} {\bibfnamefont {A.}~\bibnamefont {Koga}}, \bibinfo {author}
    {\bibfnamefont {A.~J.}\ \bibnamefont {Millis}},\ and\ \bibinfo {author}
    {\bibfnamefont {P.}~\bibnamefont {Werner}},\ }\href@noop {} {\bibfield
    {journal} {\bibinfo  {journal} {Physical Review B}\ }\textbf {\bibinfo
    {volume} {101}},\ \bibinfo {pages} {195118} (\bibinfo {year}
    {2020})}\BibitemShut {NoStop}%
  \bibitem [{\citenamefont {Murakami}\ \emph {et~al.}(2017)\citenamefont
    {Murakami}, \citenamefont {Gole{\v{z}}}, \citenamefont {Eckstein},\ and\
    \citenamefont {Werner}}]{murakami2017photoinduced}%
    \BibitemOpen
    \bibfield  {author} {\bibinfo {author} {\bibfnamefont {Y.}~\bibnamefont
    {Murakami}}, \bibinfo {author} {\bibfnamefont {D.}~\bibnamefont
    {Gole{\v{z}}}}, \bibinfo {author} {\bibfnamefont {M.}~\bibnamefont
    {Eckstein}},\ and\ \bibinfo {author} {\bibfnamefont {P.}~\bibnamefont
    {Werner}},\ }\href@noop {} {\bibfield  {journal} {\bibinfo  {journal}
    {Physical Review Letters}\ }\textbf {\bibinfo {volume} {119}},\ \bibinfo
    {pages} {247601} (\bibinfo {year} {2017})}\BibitemShut {NoStop}%
  \bibitem [{\citenamefont {Michael}\ \emph {et~al.}(2022)\citenamefont
    {Michael}, \citenamefont {Haque}, \citenamefont {Windgaetter}, \citenamefont
    {Latini}, \citenamefont {Zhang}, \citenamefont {Rubio}, \citenamefont
    {Averitt},\ and\ \citenamefont {Demler}}]{michael2022}%
    \BibitemOpen
    \bibfield  {author} {\bibinfo {author} {\bibfnamefont {M.~H.}\ \bibnamefont
    {Michael}}, \bibinfo {author} {\bibfnamefont {S.~R.~U.}\ \bibnamefont
    {Haque}}, \bibinfo {author} {\bibfnamefont {L.}~\bibnamefont {Windgaetter}},
    \bibinfo {author} {\bibfnamefont {S.}~\bibnamefont {Latini}}, \bibinfo
    {author} {\bibfnamefont {Y.}~\bibnamefont {Zhang}}, \bibinfo {author}
    {\bibfnamefont {A.}~\bibnamefont {Rubio}}, \bibinfo {author} {\bibfnamefont
    {R.~D.}\ \bibnamefont {Averitt}},\ and\ \bibinfo {author} {\bibfnamefont
    {E.}~\bibnamefont {Demler}},\ }\href@noop {} {\emph {\bibinfo {title} {Theory
    of time-crystalline behaviour mediated by phonon squeezing in Ta2NiSe5}}},\
    \bibinfo {type} {Tech. Rep.}\ (\bibinfo {year} {2022})\BibitemShut {NoStop}%
  \bibitem [{\citenamefont {Gole\ifmmode~\check{z}\else \v{z}\fi{}}\ \emph
    {et~al.}(2020)\citenamefont {Gole\ifmmode~\check{z}\else \v{z}\fi{}},
    \citenamefont {Sun}, \citenamefont {Murakami}, \citenamefont {Georges},\ and\
    \citenamefont {Millis}}]{golez2020}%
    \BibitemOpen
    \bibfield  {author} {\bibinfo {author} {\bibfnamefont {D.}~\bibnamefont
    {Gole\ifmmode~\check{z}\else \v{z}\fi{}}}, \bibinfo {author} {\bibfnamefont
    {Z.}~\bibnamefont {Sun}}, \bibinfo {author} {\bibfnamefont {Y.}~\bibnamefont
    {Murakami}}, \bibinfo {author} {\bibfnamefont {A.}~\bibnamefont {Georges}},\
    and\ \bibinfo {author} {\bibfnamefont {A.~J.}\ \bibnamefont {Millis}},\
    }\href {https://doi.org/10.1103/PhysRevLett.125.257601} {\bibfield  {journal}
    {\bibinfo  {journal} {Phys. Rev. Lett.}\ }\textbf {\bibinfo {volume} {125}},\
    \bibinfo {pages} {257601} (\bibinfo {year} {2020})}\BibitemShut {NoStop}%
  \bibitem [{\citenamefont {Tanabe}\ \emph {et~al.}(2021)\citenamefont {Tanabe},
    \citenamefont {Kaneko},\ and\ \citenamefont {Ohta}}]{tanabe2021}%
    \BibitemOpen
    \bibfield  {author} {\bibinfo {author} {\bibfnamefont {T.}~\bibnamefont
    {Tanabe}}, \bibinfo {author} {\bibfnamefont {T.}~\bibnamefont {Kaneko}},\
    and\ \bibinfo {author} {\bibfnamefont {Y.}~\bibnamefont {Ohta}},\ }\href
    {https://doi.org/10.1103/PhysRevB.104.245103} {\bibfield  {journal} {\bibinfo
     {journal} {Phys. Rev. B}\ }\textbf {\bibinfo {volume} {104}},\ \bibinfo
    {pages} {245103} (\bibinfo {year} {2021})}\BibitemShut {NoStop}%
  \bibitem [{\citenamefont {Matsunaga}\ \emph {et~al.}(2013)\citenamefont
    {Matsunaga}, \citenamefont {Hamada}, \citenamefont {Makise}, \citenamefont
    {Uzawa}, \citenamefont {Terai}, \citenamefont {Wang},\ and\ \citenamefont
    {Shimano}}]{matsunaga2013}%
    \BibitemOpen
    \bibfield  {author} {\bibinfo {author} {\bibfnamefont {R.}~\bibnamefont
    {Matsunaga}}, \bibinfo {author} {\bibfnamefont {Y.~I.}\ \bibnamefont
    {Hamada}}, \bibinfo {author} {\bibfnamefont {K.}~\bibnamefont {Makise}},
    \bibinfo {author} {\bibfnamefont {Y.}~\bibnamefont {Uzawa}}, \bibinfo
    {author} {\bibfnamefont {H.}~\bibnamefont {Terai}}, \bibinfo {author}
    {\bibfnamefont {Z.}~\bibnamefont {Wang}},\ and\ \bibinfo {author}
    {\bibfnamefont {R.}~\bibnamefont {Shimano}},\ }\href
    {https://doi.org/10.1103/PhysRevLett.111.057002} {\bibfield  {journal}
    {\bibinfo  {journal} {Phys. Rev. Lett.}\ }\textbf {\bibinfo {volume} {111}},\
    \bibinfo {pages} {057002} (\bibinfo {year} {2013})}\BibitemShut {NoStop}%
  \bibitem [{\citenamefont {Matsunaga}\ \emph {et~al.}(2014)\citenamefont
    {Matsunaga}, \citenamefont {Tsuji}, \citenamefont {Fujita}, \citenamefont
    {Sugioka}, \citenamefont {Makise}, \citenamefont {Uzawa}, \citenamefont
    {Terai}, \citenamefont {Wang}, \citenamefont {Aoki},\ and\ \citenamefont
    {Shimano}}]{matsunaga2014}%
    \BibitemOpen
    \bibfield  {author} {\bibinfo {author} {\bibfnamefont {R.}~\bibnamefont
    {Matsunaga}}, \bibinfo {author} {\bibfnamefont {N.}~\bibnamefont {Tsuji}},
    \bibinfo {author} {\bibfnamefont {H.}~\bibnamefont {Fujita}}, \bibinfo
    {author} {\bibfnamefont {A.}~\bibnamefont {Sugioka}}, \bibinfo {author}
    {\bibfnamefont {K.}~\bibnamefont {Makise}}, \bibinfo {author} {\bibfnamefont
    {Y.}~\bibnamefont {Uzawa}}, \bibinfo {author} {\bibfnamefont
    {H.}~\bibnamefont {Terai}}, \bibinfo {author} {\bibfnamefont
    {Z.}~\bibnamefont {Wang}}, \bibinfo {author} {\bibfnamefont {H.}~\bibnamefont
    {Aoki}},\ and\ \bibinfo {author} {\bibfnamefont {R.}~\bibnamefont
    {Shimano}},\ }\href@noop {} {\bibfield  {journal} {\bibinfo  {journal}
    {Science}\ }\textbf {\bibinfo {volume} {345}},\ \bibinfo {pages} {1145}
    (\bibinfo {year} {2014})}\BibitemShut {NoStop}%
  \bibitem [{\citenamefont {Tsuji}\ and\ \citenamefont {Aoki}(2015)}]{tsuji2015}%
    \BibitemOpen
    \bibfield  {author} {\bibinfo {author} {\bibfnamefont {N.}~\bibnamefont
    {Tsuji}}\ and\ \bibinfo {author} {\bibfnamefont {H.}~\bibnamefont {Aoki}},\
    }\href {https://doi.org/10.1103/PhysRevB.92.064508} {\bibfield  {journal}
    {\bibinfo  {journal} {Phys. Rev. B}\ }\textbf {\bibinfo {volume} {92}},\
    \bibinfo {pages} {064508} (\bibinfo {year} {2015})}\BibitemShut {NoStop}%
  \bibitem [{\citenamefont {Blommel}\ \emph {et~al.}(2024)\citenamefont
    {Blommel}, \citenamefont {Kaye}, \citenamefont {Murakami}, \citenamefont
    {Gull},\ and\ \citenamefont {Golež}}]{blommel2024}%
    \BibitemOpen
    \bibfield  {author} {\bibinfo {author} {\bibfnamefont {T.}~\bibnamefont
    {Blommel}}, \bibinfo {author} {\bibfnamefont {J.}~\bibnamefont {Kaye}},
    \bibinfo {author} {\bibfnamefont {Y.}~\bibnamefont {Murakami}}, \bibinfo
    {author} {\bibfnamefont {E.}~\bibnamefont {Gull}},\ and\ \bibinfo {author}
    {\bibfnamefont {D.}~\bibnamefont {Golež}},\ }\href@noop {} {\bibinfo {title}
    {Chirped amplitude mode in photo-excited superconductors}} (\bibinfo {year}
    {2024}),\ \Eprint {https://arxiv.org/abs/2403.01589} {arXiv:2403.01589
    [cond-mat.supr-con]} \BibitemShut {NoStop}%
  \bibitem [{\citenamefont {Onida}\ \emph {et~al.}(2002)\citenamefont {Onida},
    \citenamefont {Reining},\ and\ \citenamefont {Rubio}}]{onida2002}%
    \BibitemOpen
    \bibfield  {author} {\bibinfo {author} {\bibfnamefont {G.}~\bibnamefont
    {Onida}}, \bibinfo {author} {\bibfnamefont {L.}~\bibnamefont {Reining}},\
    and\ \bibinfo {author} {\bibfnamefont {A.}~\bibnamefont {Rubio}},\ }\href
    {https://doi.org/10.1103/RevModPhys.74.601} {\bibfield  {journal} {\bibinfo
    {journal} {Rev. Mod. Phys.}\ }\textbf {\bibinfo {volume} {74}},\ \bibinfo
    {pages} {601} (\bibinfo {year} {2002})}\BibitemShut {NoStop}%
  \bibitem [{\citenamefont {Kaneko}\ \emph {et~al.}(2021)\citenamefont {Kaneko},
    \citenamefont {Sun}, \citenamefont {Murakami}, \citenamefont
    {Gole\ifmmode~\check{z}\else \v{z}\fi{}},\ and\ \citenamefont
    {Millis}}]{kaneko2021}%
    \BibitemOpen
    \bibfield  {author} {\bibinfo {author} {\bibfnamefont {T.}~\bibnamefont
    {Kaneko}}, \bibinfo {author} {\bibfnamefont {Z.}~\bibnamefont {Sun}},
    \bibinfo {author} {\bibfnamefont {Y.}~\bibnamefont {Murakami}}, \bibinfo
    {author} {\bibfnamefont {D.}~\bibnamefont {Gole\ifmmode~\check{z}\else
    \v{z}\fi{}}},\ and\ \bibinfo {author} {\bibfnamefont {A.~J.}\ \bibnamefont
    {Millis}},\ }\href {https://doi.org/10.1103/PhysRevLett.127.127402}
    {\bibfield  {journal} {\bibinfo  {journal} {Phys. Rev. Lett.}\ }\textbf
    {\bibinfo {volume} {127}},\ \bibinfo {pages} {127402} (\bibinfo {year}
    {2021})}\BibitemShut {NoStop}%
  \bibitem [{\citenamefont {Shao}\ \emph {et~al.}(2016)\citenamefont {Shao},
    \citenamefont {Tohyama}, \citenamefont {Luo},\ and\ \citenamefont
    {Lu}}]{shao2016}%
    \BibitemOpen
    \bibfield  {author} {\bibinfo {author} {\bibfnamefont {C.}~\bibnamefont
    {Shao}}, \bibinfo {author} {\bibfnamefont {T.}~\bibnamefont {Tohyama}},
    \bibinfo {author} {\bibfnamefont {H.-G.}\ \bibnamefont {Luo}},\ and\ \bibinfo
    {author} {\bibfnamefont {H.}~\bibnamefont {Lu}},\ }\href
    {https://doi.org/10.1103/PhysRevB.93.195144} {\bibfield  {journal} {\bibinfo
    {journal} {Phys. Rev. B}\ }\textbf {\bibinfo {volume} {93}},\ \bibinfo
    {pages} {195144} (\bibinfo {year} {2016})}\BibitemShut {NoStop}%
  \end{thebibliography}
\end{document}